\documentclass[11pt]{article}
\usepackage{graphicx,amsmath,amsfonts,changebar,color}
\usepackage{color}
\newcommand{\MP}[1]{\marginpar {\tiny {}}}

\newenvironment{Proof}{\noindent{\bf Proof:\ }}{
                        \vspace{0.25cm}\par}
\newenvironment{Prooflemma}{\noindent{\bf Proof:\ }}{
                       \vspace{0.25cm}\par}              
\newtheorem{proposition}{Proposition}[section]
\newtheorem{theorem}[proposition]{Theorem}

\newtheorem{lemma}[proposition]{Lemma}

\newtheorem{corollary}[proposition]{Corollary}
\newtheorem{definition}[proposition]{Definition}
\newcommand{\0}{\vec{\mathbf 0}}
\newcommand{\1}{\vec{\mathbf 1}}

\newcommand{\ren}{\stackrel{\sim}{\rightarrow}}

\newcommand{\Cons}{\mathrm{Cons}}
\newcommand{\WAg}{\mathrm{WAg}}
\newcommand{\AC}{\mathrm{AC}}
\newcommand{\kT}{$k$-$\mathrm{TAg}$}
\newcommand{\lT}{$(k\!-\!1)$-$\mathrm{T\!Ag}$}
\newcommand{\ukT}{$(k\!+\!1)$-$\mathrm{T\!Ag}$}
\newcommand{\fT}{$f$-$\mathrm{T\!Ag}$}
\newcommand{\ufT}{$(f\!+\!1)$-$\mathrm{T\!Ag}$}
\newcommand{\N}{\mbox{\rm I\hspace{-.5em}N}}

\newcommand{\send}{\mbox{\em Send\/}}
\newcommand{\query}{\mbox{\em Query\/}}
\newcommand{\answ}{\mbox{\em Answer\/}}
\newcommand{\rec}{\mbox{\em Receive\/}}
\newcommand{\dec}{\mbox{\em Decide\/}}

\newcommand{\C}{{\,\leq_C\,}}
\newcommand{\FI}{{\,\leq_{FI}\,}}
\newcommand{\SC}{{\,\leq_{C^*}\,}}

\newcommand{\qedsymbol}{$\Box$}
\newcommand{\qed}{\nopagebreak \hfill \qedsymbol \nopagebreak}

\begin{document}

\title{Reductions in Distributed Computing\\ 
\vspace{0.3cm}
\Large{Part II: $k$-\!Threshold Agreement Tasks} } 
\author{Bernadette Charron-Bost\thanks{Laboratoire LIX, 
	\'Ecole Polytechnique, 91128 	
	Palaiseau Cedex, France}}

\date{\empty} 
\maketitle

\begin{abstract}
We extend the results of Part~I by considering a new  class of agreement tasks,
	the so-called $k$-Threshold Agreement tasks (previously introduced by Charron-Bost and Le 		Fessant).
These tasks naturally interpolate between Atomic Commitment and Consensus.
Moreover, they constitute a valuable tool to derive irreducibility results between Consensus tasks
	only.
In particular, they allow us to show that  (A) for a fixed set of processes, the higher the resiliency 
	degree is,  the harder the Consensus task is, 
	and (B)  for a fixed resiliency degree, the smaller the set of processes is,  the harder the 
	Consensus task is.

The proofs of these results  lead us to consider new oracle-based reductions, involving 
	a weaker variant of the $C$-reduction introduced in Part~I.
We also discuss the relationship  between our results and previous ones relating 
	  $f$-resiliency and wait-freedom.
	
\end{abstract}

\section{Introduction}

In Part~I of  this paper, we developed several formal definitions of reduction in distributed computing 
	that allowed us to formalize in which sense some distributed task is easier to solve than another 
	one.
We applied this formalism for reduction to compare two fundamental classes of agreement tasks, 
	namely  {\em Binary Consensus} and {\em Atomic Commitment}: we  showed that even if 			Consensus and  Atomic Commitment are syntactically very close, these two types of task are 			incomparable 
	in most cases in the sense that  Consensus is not reducible to  Atomic Commitment, and 
	{\it vice-versa}.\footnote{More precisely, Consensus and Atomic Commitment are not comparable,
	except when the resiliency degree is 1 in which case Consensus is easier than Atomic 
	Commitment.} 
	
Here in Part~II, we consider the new class of agreement tasks introduced by Charron-Bost 
	and Le Fessant~\cite{CL04}, the so-called $k$-{\em Threshold Agreement} tasks 
	($k$-TAg, for short).
Those tasks interpolate between Atomic Commitment and Consensus from a purely syntactic 				standpoint: for the lowest parameter value $k=1$, $k$-TAg coincides with Atomic Commitment,
	and for the highest parameter value $k=n$ (where $n$ is the number of processes) $k$-TAg 			coincides with Consensus (see Section~2 in~\cite{CL04} and {\it infra}).
	
We begin by comparing the various agreement tasks $k$-TAg$(n,f)$ when varying the parameter $k$, 
	the number of processes $n$, and the resiliency degree $f$. 
We do that in  generalizing each of the reducibility and irreducibility results established in Part~I for 
	Atomic Commitment and Consensus, and in extending their scope to the general class of 
	$k$-Threshold Agreement tasks.
	
Then, by combining these results for the $k$-TAg tasks, 
	we derive new irreducibility results for  Consensus tasks only. 
Notably,  we establish two irreducibility results both revealing that ``wait-freedom is harder to achieve 
	than $f$-resiliency'' in the case of Consensus tasks.
More precisely, we show that (A) for a fixed set of processes, the higher the resiliency degree is, 
	the harder the Consensus task is, and (B)  for a fixed resiliency degree, the smaller the set of 			processes is,  the harder the Consensus task is.

When the resiliency degree $f$ is less than $n/2$ (that is, when a majority of processes is correct), 
	the fact that wait-free Consensus is harder than $f$-resilient Consensus  is not very surprising.
Indeed, Fischer, Lynch, and Paterson~\cite{FLP85} have shown that  in the more benign failure
	 model of initial crashes,  $f$-resilient Consensus is solvable with a minority of faulty processes, 		whereas wait-free Consensus is not solvable with initial crashes.
However, our two irreducibility results (A) and (B) are more surprising in view of prior work
	comparing Consensus tasks in  the message passing and  shared memory models.

Firstly, an immediate corollary of the main results established by Chandra, Hadzilacos, and 
	Toueg~\cite{CT96,CHT96} is that for a fixed set of processes, the same information about failures
	is necessary and sufficient to solve all the Consensus tasks with a majority of correct processes.
Using the notion of {\em Failure Detectors}, this can be rephrased by saying that the weakest 
	Failure Detectors to solve the Consensus tasks with a majority of correct processes are identical.
From this viewpoint, all these tasks are therefore equivalently hard to solve.
On the contrary, we show that they are not equivalent with respect to any of the reductions 
	defined in Part~I -- in particular with respect to the most natural and meaningful reduction 
	in this context, namely the $C$-reduction.
Compared with (A), the results in~\cite{CT96, CHT96} show that, if we introduce the 
	{\em Failure Information hierarchy} (FI-hierarchy, for short) which assesses the hardness to 
	solve a task only by  the minimal information  about failures that is required to solve it, then the 		$C$-hierarchy  is strictly finer than the  FI-hierarchy.
In other words, the  minimal information about failures -- or equivalently  the weakest Failure Detector -- 		necessary for  solving a task does not fully capture the hardness to solve it.

As for our second irreducibility result (B), it seems to contradict
	Borowsky and Gafni's simulation~\cite{BG93}, and more specifically its 
	variants described by Lo and Hadzilacos~\cite{LH00} and 
	by Chandra, Hadzilacos, Jayanti, and Toueg~\cite{CHJT94} 
	  in the case of Consensus tasks.
Indeed, this simulation consists in a general algorithm for the shared memory model which allows 
	a set of $f+1$ processes with at most $f$ crash failures to simulate any larger set of $n$ 
	processes also with at most $f$ crashes.
In the case of Consensus tasks, this simulation provides a general transformation of algorithms 
	that solve the $f$-resilient Consensus task for $n$ processes using read/write registers into 			ones that solve the wait-free Consensus task for $f+1$ processes using read/write registers
	also.
On the contrary, we show that in the message passing model, the wait-free Consensus task for 
	$f+1$ processes is not $C$-reducible to the $f$-resilient Consensus task for a strict superset of 
	$n$ processes.
We could think that the discrepancy between the work in the papers cited  above and our 
	results comes from  the fact that the message passing and the shared memory models are
	 precisely not  equivalent here (a majority of processes may fail in the wait-free case).
However,  a closer look at the transformations of algorithms reveals  that the discrepancy actually 
	results from a more basic point.
Indeed, in our work, we use notions of reductions which rely on suitably defined 
	{\em distributed oracles}, which are closed black boxes that cannot be opened.
We have no access to their internal mechanisms, and are not allowed to dismantle them 
	and to distribute them 
	onto the different processes in the system.
This is the reason why the transformations in~\cite{CHJT94,LH00} have no translation in terms of 
	oracle-based reductions.
So the discrepancy between these prior results and ours  reflects substancial differences 				between the underlying types of reduction.
This highlights the need to give precise definitions to the reductions that we handle.

Another contribution of Part~II is to introduce new oracle-based reductions.
Namely, when comparing the hardness to solve the two $f$-resilient tasks $f$-TAg$(n,f)$ 
	and $(f\!+\!1)$-TAg$(n,f)$, we have been led to consider various weakenings of the 
	$C$-reduction in which the oracles are more powerful than those used for the $C$-reduction.
The resulting  new kinds of reductions differ from the $C$-reduction  not in the way processes query 		oracles, but rather  in the quality of the oracles that are used.
More precisely, to each task $T=(P,f)$ we associate some $f$-resilient oracle suitable for $P$
	which is more deterministic  than the oracle ${\cal O}.T$ defined in Part~I, in the sense that 
	the set of all its possible behaviors is smaller.
Such an  oracle is  therefore more powerful than ${\cal O}.T$, and so yields a weaker type of 
	reduction {\em \`a la} Cook.

Part~II  of this paper  is organized as follows.
In Section~2, we introduce the $k$-Threshold Agreement tasks, and derive some simple reductions
	between these tasks  from their specifications only.
Then, in Sections~3 and~4,  we generalize the $C$- and $C^*$-reducibility results of  Part~I
	to this new class of agreement tasks. 
In Section~5, we give generalizations of the irreducibility statements established in Part~I, 
	completing the picture of the $C$-hierarchy between the $k$-Threshold Agreement tasks.
Then we derive irreducibility results between Consensus tasks only, and we show that wait-free
	Consensus is strictly harder to achieve than $f$-resilient Consensus in Section~6.
This section proceeds with a discussion of these results comparing them  with prior work  relating 
	wait-freedom and $f$-resiliency.

\section{$\mathbf{k}$-Threshold Agreement tasks}

\subsection{Definitions and notation}

The main results in Part~I show that $f$-resilient Consensus and Atomic Commitment tasks are 
	generally not comparable from an algorithmical point of view:  except in the case $f=1$, 
	there is no algorithm which converts a solution to Consensus into a solution to Atomic 
	Commitment, and {\it vice-versa}.
	
However, Consensus and Atomic Commitment problems are very close in the sense that 
	their specifications are  identical, except the validity conditions which are slightly 
	different.
As a matter of fact, it is possible to link  these two specifications from a purely syntactic
	standpoint, as was done in~\cite{CL04} by Charron-Bost and Le Fessant who introduced, 
	for any set $\Pi$ of $n$ processes and any integer $k\in\{1,\cdots,n\}$,
	the $k$-{\em Threshold Agreement} 	problem for $\Pi$ (the $k$-TAg$_{\Pi}$ problem, for short).
Parameter $k$ is called the {\em threshold value} of   $k$-TAg$_{\Pi}$.

Formally,  we have ${\cal V}=\{0,1\}$, ${\cal V}_{k\mbox{-TAg}_{\Pi}} =\{0,1\}^{\Pi}$, 
	 and for any  $(F,\vec{V})\in {\cal F}_{\Pi}\times \{0,1\}^{\Pi}$,
\begin{itemize}
\item \kT$_{\Pi}(F,\vec{V})= \{ 0 \} \mbox{ if } |\{p\in \Pi : \vec{V}(p)=0\}|\geq k$,
\item \kT$_{\Pi}(F,\vec{V})= \{ 1 \} \mbox{ if } \vec{V} = \1 \mbox{ and }
	|Faulty(F)| \leq k-1$,
\item \kT$_{\Pi}(F,\vec{V})= \{ 0,1 \} \mbox{ otherwise.}$
\end{itemize}
In other words, the  $k$-TAg validity condition expresses the fact that
	(1)  if at least $k$ processes start with 0, then 0 is the only possible
	decision value, and (2) if all processes start with 1 and  at most $k-1$  failures occur,
	then 1 is the only possible decision value.

This new problem is a straightforward generalization of the Atomic Commitment
	problem (1-TAg coincides with Atomic Commitment), but turns out to be also a
	generalization of Consensus.
Indeed, $n$-TAg actually corresponds to binary Consensus,
	since at least one process in $\Pi$ is correct.
We have thereby defined a chain of problems which interpolates between
	 Atomic Commitment and (binary) Consensus.
The main motivation for this generalization is theoretical: it is
	interesting to connect two incomparable problems by exploiting
	differences in validity conditions.
But it is easy to imagine actual situations in which such a generalization
	arises naturally: for example, it might be desirable for processes
	to enforce them to decide 0 ({\tt abort}) only if a majority of processes 
	initially propose 0 ({\tt no}),
	but to require that they decide 1 (commit) if all of them initially propose 1
	({\tt yes}) as soon as a minority of processes are faulty.
Indeed, the latter problem corresponds exactly to the $k$-TAg problem with the threshold value
	$k=\lfloor n/2\rfloor +1$.

For any integer  $f$ such that $0\leq f\leq n-1$, we denote  \kT$(\Pi,f)$ the distributed task 
	defined by the  problem $k$-TAg$_{\Pi}$ and the resiliency degree $f$.
Hence we have a chain of $n$ tasks 1-TAg$(\Pi,f),\cdots,n$-TAg$(\Pi,f)$ which syntactically 
	relates $\AC(\Pi,f)= 1\mbox{-TAg}(\Pi,f)$ to $\Cons(\Pi,f)= n\mbox{-TAg}(\Pi,f)$.
Observe that two consecutive tasks in this chain are not comparable {\it a priori},
	since the two parts of the validity condition are entangled:
	the first part  for $k$-TAg$_{\Pi}$ enforces the first part for $(\! k\!+\!1\! )$-TAg$_{\Pi}$,
	and  the second part for $k$-TAg$_{\Pi}$  is implied by  the second part for  
	$(\! k\!+\!1\! )$-TAg$_{\Pi}$.
	
Recall that Fischer, Lynch, and Paterson~\cite{FLP85} have established the impossibility of 
	some agreement task with resiliency degree 1 that is attached to a problem weaker
	in its validity condition than all the above $k$-TAg problems.\footnote{Namely, the validity 
	condition in~\cite{FLP85} only specifies that for every value $v\in {\cal V}$, there is an execution 
	in which some process decides $v$.}
	
Consequently, for any set $\Pi$ of $n$ processes and  for any 
	integers $k$ and $f$ such that $1\leq k \leq n$ and $1\leq f \leq n-1$,  the task \kT$(\Pi,f)$
         is not solvable in an asynchronous system.
	
In the sequel, we denote by \kT$(n,f)$ the $f$-resilient task defined 
	by the $k$-Threshold Agreement problem for the set of process names
	$\Pi=\{1,\cdots, n\}$, that is:	
	$$k\mbox{-TAg}(n,f) =  k\mbox{-TAg} \left  ( \{1,\cdots, n\},f \right ).$$
Clearly, for any set $\Pi$ of $n$ processes and for any renaming $\Phi : \{1,\cdots, n\} \ren \Pi$, 
	we have:
\begin{equation}\label{kTAgn}
^{\Phi} k\mbox{-TAg}(n,f) = k\mbox{-TAg} (\Pi, f),
\end{equation}
	and consequently like  $\Cons(\Pi,f)$ and  $\AC(\Pi,f)$, \kT$(\Pi,f)$ is a symmetric task.
	
\subsection{Oracles suitable for $\mathbf{k}$-TAg problems}\label{Ora}

Following Section~I.3.3, to each task \kT$(\Pi,f)$, we associate a unique oracle,
	denoted ${\cal O}$.\kT$(\Pi,f)$, which is the most general $f$-resilient oracle suitable
	for the agreement problem $k$-TAg$_{\Pi}$.

Several results in the sequel rely on the following claims about the ${\cal O}$.\kT$(\Pi,f)$
	oracles.\\

\noindent{\large {\bf Claim O1:}}
{\em For any failure pattern $F$ for $\Pi$, and for any consultation in a history of 
	 ${\cal O}$.\kT$(\Pi,f)(F)$, 
	if at most $|\Pi|-k$ queries have value 1 (that is at least $k$ queries are either missing or 
	have value 0), then the only possible response of the oracle is 0.}\\

\begin{Proof}
Form the partial vector $\vec{W}$ with the queries in the consultation of ${\cal O}$.\kT$(\Pi,f)$.
Since at most $|\Pi|-k$ components of $\vec{W}$ are equal to 1, there is an extension $\vec{V}$
	of $\vec{W}$ in $\{0,1\}^{\Pi}$ with at least $k$ components equal to 0.
By the first part in the validity condition of $k$-TAg$_{\Pi}$, we have 
	$k$-TAg$_{\Pi}(F,\vec{V}) = \{0\}.$
Hence, 0 is the only possible response by  ${\cal O}$.\kT$(\Pi,f)(F)$ in the consultation with the
	query vector $\vec{W}$.
\qed
\end{Proof}

\noindent{\large {\bf Claim O2:}}
{\em For any failure pattern $F$ for $\Pi$ with less than $k$ faulty processes, and for any consultation 
	in a history of  ${\cal O}$.\kT$(\Pi,f)(F)$, 
	if all the query values are 1, then the only possible response of the oracle is 1.}\\

\begin{Proof}
This is a straightforward consequence of the second part in the validity condition of 
	$k$-TAg$_{\Pi}$.\qed
\end{Proof}

Notice that  the property O$_{\Cons}$ of Consensus oracles (cf. Section~I.3.3) 
	coincides with the conjunction of Claim O1 and Claim O2 for  the threshold value $k=|\Pi|$,
	and  the property O$_{\AC}$ of Atomic Commitment oracles (cf. Section~I.3.3, too) 			coincides with Claim O1 for the threshold value $k=1$.

\subsection{Some $\mathbf{k}$-TAg tasks are generalizations of Consensus}

One trivial but useful kind of $K$-reduction is {\em reduction by generalization}.
We say that task $T_2$ is a {\em generalization} of task $T_1$ when, informally, the resiliency 
	degree of $T_2$ is not greater than the one of $T_1$,  the inputs for $T_1$ are inputs 
	for $T_2$, and for those inputs, any solution for $T_2$ is  also a solution for $T_1$.
Formally, 	$T_1=(P_1,f_1)$ is a generalization of $T_2=(P_2,f_2)$ if $f_1\leq f_2$, 
	${\cal V}_{P_1}\subseteq  {\cal V}_{P_2}$, and for  any input vector 
	$\vec{V}\in {\cal V}_{P_1}$ and any failure pattern $F$ such that $|Faulty(F)|\leq f_1$,
	 we have $P_2(F,\vec{V})\subseteq P_1(F,\vec{V})$.
When these conditions are satisfied, we shall also say  that $T_1$ is a {\em special case} of $T_2$.
Notice that if $T_2$ is a generalization of $T_1$, then there is a trivial $K$-reduction from 
	$T_1$ to $T_2$: just take $R$ to be the algorithm which does nothing.
	
A first example of reduction by generalization is the reduction from any task 
	$T_1=(P,f_1)$ to $T_2=(P,f_2)$ with $f_1\leq f_2$.
To illustrate this notion with a less trivial example, consider the {\em Weak Agreement}  problem 			for $\Pi$ introduced in~\cite{Lam83b}, denoted $\WAg_{\Pi}$ and defined 
	 by ${\cal V}_{\WAg_{\Pi}} = \{0,1\}^{\Pi}$ and for any 
	$(F,\vec{V})\in {\cal F}_{\Pi}\times \{0,1\}^{\Pi}$,
\begin{itemize}
\item $\WAg_{\Pi}(F,\vec{V})= \{ 0  \} \mbox{ if } \vec{V} = \0 \mbox{ and } Faulty(F)=\emptyset$;
\item $\WAg_{\Pi}(F,\vec{V})= \{ 1 \} \mbox{ if } \vec{V} = \1 \mbox{ and } Faulty(F)=\emptyset$;
\item $\WAg_{\Pi}(F,\vec{V})= \{ 0,1 \}$,  otherwise. 
\end{itemize}
Both $\Cons(\Pi,f)$ and $\AC(\Pi,f)$ are generalizations of 
	$\WAg(\Pi,f)=(\WAg_{\Pi},f)$.
More generally, $\WAg(\Pi,f)$ is a special case of  \kT$(\Pi,f)$ for any $k\in\{1,\ldots,n\}$. 

As we shall show in Proposition~\ref{generalization}, there is a chain of reductions by 
	generalization that can be traced among the set of $f$-resilient tasks 
	$\{1$-TAg$(n,f), \cdots, n$-TAg$(n,f)\}$ from threshold $k=f\!+\!1$.
	
\begin{proposition}\label{generalization}
If $n$ and $f$  are  two integers such that $1\leq f  \leq  n-1$, 
	then for any $k\in\{f\!+\!1,\cdots, n\!-\!1\}$, the task \kT$(n,f)$  is 
	a generalization of  \ukT$(n,f)$, and so \ufT$(n,f)$ is a generalization of $\Cons(n,f)$.
\end{proposition}

\begin{Proof}
Let $\Pi$ be any set of $n$ processes.
We only have to prove that for  any input vector 
	$\vec{V}\in \{0,1\}^{\{1,\cdots,n\}}$ and any failure pattern $F$ for $\Pi$ 
	such that $|Faulty(F)|\leq f$,  we have 
	 $$k\mbox{-TAg}_{\Pi}(F,\vec{V})\subseteq (k\!+\!1)\mbox{-TAg}_{\Pi}(F,\vec{V}).$$
This inclusion is obvious when \ukT$_{\Pi}(F,\vec{V}) =\{0,1\}$.
Therefore, we need to consider the following two non-trivial cases only:
\begin{enumerate}
\item $|\{p\in \Pi : \vec{V}(p)=0\}|\geq k\!+\!1$.\\
{\it A fortiori}, $|\{p\in \Pi : \vec{V}(p)=0\}|\geq k$ and thus we have 
	 $$k\mbox{-TAg}_{\Pi}(F,\vec{V}) =  (k\!+\!1)\mbox{-TAg}_{\Pi}(F,\vec{V}) = \{ 0 \}.$$	
\item $\vec{V}=\1$ and $|Faulty(F)| \leq k$.\\
Since we only examine the failure patterns with at most $f$ failures and $k\geq f\!+\!1$,
	we actually have $|Faulty(F)| \leq k-1$.
It follows that  
	 $$k\mbox{-TAg}_{\Pi}(F,\1) =  (k\!+\!1)\mbox{-TAg}_{\Pi}(F,\1) = \{ 1 \}.$$	 
\end{enumerate}\qed
\end{Proof}

\section{C-reductions between $\mathbf{k}$-Threshold Agreement tasks}

This section is devoted to several generalizations of the $C$-reductions established in Part~I.
First we shall show that \kT$(n,f)$ is $C$-reducible to \ukT$(n\!+\!1,f\!+\!1)$.
Then we shall complete Proposition~\ref{generalization} and the comparison of the 
	tasks \ufT$(n,f), \ldots, n\mbox{-TAg}(n,f)$ by showing  that all these tasks are actually
	equivalent --  that is,  of the same unsolvability degree -- with respect to the $C$-reduction.
Finally, we shall consider \kT$(n,f)$ when the threshold value $k$ takes the preceding value $f$, 
	and shall compare the task $f\mbox{-TAg}(n,f)$ with $n\mbox{-TAg}(n,f)=\Cons(n,f)$.
For that, we shall introduce a slightly weaker notion of reduction {\em \`a la Cook}, 
	the $^cC$-reduction, which differs from the original $C$-reduction in the power of oracles
	(but not in the way oracles are queried).
We shall show that when a majority of processes is correct, i.e., when $n>2f$,
	the degree of unsolvability --~with respect to the  $^cC$-reduction~-- of \fT$(n,f)$ is  higher or equal 		to the one of $\Cons(n,f)$.

\subsection{$\mathbf{C}$-reduction between $\mathbf{k}$- and $\mathbf{(k\!+\!1)}$-TAg tasks}

Let $\Pi$ be any set of $n+1$ processes, and let $\Pi' $ be any subset of $ \Pi$ 
	with  $n$ processes.
The $C$-reduction from \kT$(\Pi',f)$ to \ukT$(\Pi,f\!+\!1)$ is simple:
Each process in $\Pi'$ just needs to query the oracle ${\cal O}.$\ukT$(\Pi,f\!+\!1)$ with its initial value.
The oracle eventually gives a response since it is consulted by at least $n-f=(n\!+\!1)-(f\!+\!1)$ 				processes.
Every process finally decides on the value provided by ${\cal O}.$\ukT$(\Pi,f\!+\!1)$.

Thus termination and agreement are obviously  guaranteed.
For the $k$-validity condition, consider a run of this algorithm for $\Pi'$ with at most $f$ failures.
Firstly, suppose that  at least $k$ processes start with 0; at least $k+1$ queries in the consultation
	of  ${\cal O}.$\ukT$(\Pi,f\!+\!1)$ are either missing or with value 0 since the process in 
	$\Pi\setminus\Pi'$ does not query the oracle.
By Claim~O1, the oracle may not answer any value else than 0, and so the only possible 
	decision value is 0.
Now, suppose that  all processes start with 1 and at most $k-1$ processes fail in this run.
With respect to $\Pi$,  at most $k$ processes crash, and by Claim~O2 applied to the oracle  
	${\cal O}.$\ukT$(\Pi,f\!+\!1)$, the only possible 
	answer value given by this oracle is 1.
Therefore, 1 is the only possible decision value.
This establishes:
	
\begin{theorem}\label{+1+1+1}
If $n, f$ and $k$  are  three integers such that $1\leq f  \leq  n-1$ and $1\leq k  \leq  n$, 
	then \kT$(n,f)$ is $C$-reducible to \ukT$(n\!+\!1,f\!+\!1)$.
\end{theorem}

Notice that in the particular case $k=n$, Theorem~\ref{+1+1+1} states that $\Cons(n,f)$ is
	$C$-reducible to $\Cons(n\!+\!1,f\!+\!1)$, and consequently it generalizes Proposition~I.7.5. 
	
\subsection{Degree of unsolvability of Consensus tasks}

As stated in  Proposition~\ref{generalization},  each task \kT$(n,f)$ with $k\in \{f\!+\!1,\ldots, n\}$
	is a generalization of $\Cons(n,f)$.
Hence $\Cons(n,f)$ trivially $C$-reduces to any of these $k$-Threshold Agreement  tasks.
We shall next show that conversely \ufT$(n,f)$ is $C$-reducible to $\Cons(n,f)$.
It will then follow that any \kT$(n,f)$ with $k\in \{f\!+\!1,\ldots, n\}$ is equivalent to $\Cons(n,f)$ with
	respect to $C$-reducibility.
In other words, \ufT$(n,f), \ldots, n$-TAg$(n,f)=\Cons(n,f)$ have the same unsolvability degree.

Let $\Pi$ be any set of $n$ processes.
There is quite a simple $C$-reduction from  \ufT$(\Pi,f)$ to $\Cons(\Pi,f)$:
Firstly, processes make their initial values more uniform.
For that, every process sends its initial value to all, waits until receiving initial values from  $n-f$ 			processes, and then sets a local variable to the minimum 
	value that it has received.
Secondly,  each process queries the oracle ${\cal O}.\Cons(\Pi,f)$ with the value of this local variable,
	and then decides on the value answered by the oracle.

\begin{figure*}[t]\small
$\hrulefill$
\begin{tabbing}
mm\=mm\=mm\=mm\=mm\=mm\=mm\=mm\=mm\=mm\=mm\= \kill
\textbf{Variables of process} $p:$\\  \\
\> $x_p\in V$, initially $v_p$\\ \\
\textbf{Algorithm for  process} $p:$\\  \\
\> $\send \langle v_p \rangle$ to all\\
\>\textbf{wait until} [$\rec \langle v_q \rangle $ from  $n-f$ processes]\\
\>  $x_p:=\min \{v_q\ :\ \mbox{received } v_q\}$\\
\> $\query ({\cal O}.\Cons(\Pi,f)) \langle x_p \rangle $\\
\> $\answ({\cal O}.\Cons(\Pi,f)) \langle d\rangle $\\
\> $\dec(d)$\\
\end{tabbing}
\vspace{-0.5cm}
\caption{A $C$-reduction from \ufT$(\Pi,f)$ to $\Cons(\Pi,f)$.}
\label{figdegcons}
$\hrulefill$
\end{figure*}

\begin{proposition}\label{eqcons}
Let $n$ and $ f$  be  two integers such that $1\leq f  \leq  n-1$, and let $\Pi$ be a set of $n$
	processes.
The algorithm in Figure~\ref{figdegcons}, that  uses the oracle $\Cons(\Pi,f)$,
	solves the task  ${\cal O}$.\ufT$(\Pi ,f)$, and so 
	\ufT$(n,f)$ is $C$-reducible to $\Cons(n,f)$.
\end{proposition}

\begin{Proof}
Let  $\rho=<\!\! F, I, H \!\!>$ denote a run of the algorithm in Figure~\ref{figdegcons} with at most 
	$f$ failures.
Obviously, $\rho$  satisfies the termination, irrevocability, and agreement conditions.
We are going to prove that $\rho$ also satisfies the two requirements of $(f\!+\!1)$-validity condition.
\begin{enumerate}
\item Suppose that  at least $f\!+\!1$ processes start with value 0.
Since at most $f$ processes fail,  each process that is still alive receives 
	at least one message with value 0, and so queries the ${\cal O}.\Cons(\Pi,f)$ oracle with value 0.
From the property O$_{\Cons}$ in Part~I, it follows that ${\cal O}.\Cons(\Pi,f)$  
	definitely answers 0.
Therefore every process that makes a decision decides 0.

\item  Now suppose that all the processes start with value 1, all the query values of 
	${\cal O}.\Cons(\Pi,f)$ are equal to 1.
Again by the property O$_{\Cons}$,  the only possible answer given by  
	${\cal O}.\Cons(n,f)$  is 1, and processes decide 1.
\end{enumerate}\qed
\end{Proof}

\subsection{A reduction with a majority of correct processes}\label{consred}

Propositions~\ref{generalization} and \ref{eqcons} establish that \ufT$(n,f), \ldots, n$-TAg$(n,f) =
	\Cons(n,f)$ are all of same degree of unsolvability with respect to  $\C$.
In this section, we compare these equivalent tasks with  \fT$(n,f)$:  we show that 
	if $n>2f$, then  \fT$(n,f)$ is at least as hard to solve as  $\Cons(n,f)$. 

In the case $f=1$, this result will compare $\Cons(n,1)$ with	 $\AC(n,1)$ when $n>2$.
However, the reduction that we shall describe does not coincide with the reduction from $\Cons(n,1)$
	to $\AC(n,1)$ given in Proposition~I.8.4.
Indeed, contrary to this latter reduction, our algorithm which uses an  oracle  for \fT$(n,f)$ actually 
	solves $\Cons(n,f)$ only if  the oracle  is {\em consistent}, namely if it satisfies the following
	condition:  if the oracle answers 0 in some consultation in which all the query values are 
	equal to 1, then it will also answer 0 in any subsequent consultation.
In other words, 	a consistent oracle for  \fT$(n,f)$ which answers some value on the grounds of 
	informations about future failures, does not forget these informations and takes them into 
	account in its subsequent answers as it does in its previous answers.
	
We are going to define consistent oracles precisely.
First, let us observe the following fact: if  $P$ is an agreement problem for  some set 
	$\Pi$  of processes, and if $\vec{W_1}$ and  $\vec{W_2}$ are two partial vectors in ${\cal V}^{\Pi}$
	such that $\vec{W_1}$ is an extension of $\vec{W_2}$ (denoted $\vec{W_1}\geq \vec{W_2}$),
	\footnote{Recall that  $\vec{W_1}$ is an {\em extension} of $\vec{W_2}$ if the domain of definition 
	$\Pi_1$ of the mapping $\vec{W_1}$ contains the one $\Pi_2$ of  $\vec{W_2}$, 
	and for any $p\in\Pi_2$, we have  $\vec{W_1}(p) = \vec{W_2}(p)$.} 
	 then for any failure pattern $F$ for $\Pi$ we have 
	$$\bigcap_{\{\vec{V}\in{\cal V}^{\Pi}\ :\ \vec{V} \geq \vec{W_2}\}} P(F,\vec{V})\subseteq 
	\bigcap_{\{ \vec{V}\in {\cal V}^{\Pi}\ :\ \vec{V} \geq \vec{W_1}\}} P(F,\vec{V}).$$                                                                
\begin{definition}
Let  ${\cal O}_{\sigma}$ be an oracle whose set of consultants is $\Pi$, and which is suitable 
	for some agreement problem $P$ for $\Pi$.
We say that ${\cal O}_{\sigma}$ is  a {\em consistent  oracle} if for any failure pattern $F$ 
	for $\Pi$,  any history $H \in{\cal O}_{\sigma}(F)$, and any two  consultations
	of ${\cal O}_{\sigma}$ in $H$ with  query vectors $\vec{W_1}$ and $\vec{W_2}$ such that 
	 $\vec{W_1}\geq \vec{W_2}$, ${\cal O}_{\sigma}$  answers $d$ in the  consultation with 
	 query vector $\vec{W_2}$ only 
	 if it answers $d$ in the consultation  with 
	 query vector $\vec{W_1}$.
\end{definition}

For any task $T=(P,f)$, we restrain the set of histories of  the oracle  in order
	to get the most general oracle which is consistent, $f$-resilient, and suitable for $P$.
In this way, we obtain an oracle, denoted $^c{\cal O}.T$, which is at least as powerful as ${\cal O}.T$
	in the sense that for any failure pattern $F$ for $\Pi$,
	$^c{\cal O}.T(F)\subseteq {\cal O}.T(F)$.

This yields a new notion of reduction {\em \`a la Cook}, denoted $\leq_{^cC}$,  in which algorithms 
	may only use the consistent versions of oracles.
Formally, $T_1\leq_{^cC} T_2$ if there is an algorithm
	for $T_1$ using the consistent  oracle $^c{\cal O}.T_2$.
	
Since $^c{\cal O}.T$ is at least as powerful as  ${\cal O}.T$, $C$-reducibility implies 
	$^cC$-reducibility.	
Equivalently, $^cC$-irreducibility results yield the corresponding $C$-irreducibility results
	(see Section~\ref{waitfres} {\it infra}).
	
Thanks to this new notion of reducibility, we shall be able to compare the two tasks  \fT$(\Pi,f)$ 
	and  \ufT$(\Pi,f)$.

Let $\Pi$ be any set of $n$ processes, and let $f$ be a positive  integer such that $n>2f$.
In Figure~\ref{figfnf}, we give an algorithm using the consistent oracle  for  \fT$(\Pi,f)$ which solves 			the task \ufT$(\Pi,f)$.
Our algorithm uses the oracle $^c{\cal O}.$\fT$(\Pi,f)$ twice:  the first time  to achieve an 
	``approximate $(f\!+\!1)$-Threshold Agreement'' on the initial values, and the second time  
	to get some informations 	about failures.
More precisely, with the help of this oracle, processes 
	determine whether less than $f$ failures occur or not in the run.
In the first case (less than $f$ failures), a solution for the $f$-TAg$_{\Pi}$  problem is also 
	a solution for  \ufT$_{\Pi}$.
If exactly $f$ failures occur, the algorithm is designed so that processes never make a wrong 				decision, and  correct processes make a decision at the latest just after the last crash.
The complete code of the $^c C$-reduction is given in Figure~\ref{figfnf}.

\begin{figure*}\small 
$\hrulefill$
\begin{tabbing}
mm\=mm\=mm\=mm\=mm\=mm\=mm\=mm\=mm\=mm\=mm\= \kill
\textbf{Variables of process} $p:$\\ \\
\> $x_p\in V$, initially $v_p$\\
\> $r_p\in \N$, initially 1\\ \\
\textbf{Algorithm for  process} $p:$\\ \\
\> $\query (^c{\cal O}.$\fT$(\Pi,f)) \langle v_p \rangle $\\
\> $\answ (^c{\cal O}.$\fT$(\Pi,f)) \langle v \rangle $\\
\> $\query(^c{\cal O}$.\fT$(\Pi,f)) \langle 1  \rangle $\\
\> $\answ(^c{\cal O}$.\fT$(\Pi,f)) \langle d \rangle $\\
\>\textbf{if} $d=1$\\
\>\textbf{then}\\
\>\> $\dec(v)$\\
\>\textbf{else}\\
\>\> \textbf{repeat forever} \\
\>\> \> $\send  \langle (R,x_p,r_p) \rangle $ to all\\
\>\>\> \textbf{wait until} [$\rec \langle (R,*,r_p) \rangle $ from $n-f$ processes] 
                                      (where $*$ can be 0 or 1)\\
\>\>\> \textbf{if} at least  $f\!+\!1$ of the  $\langle (R,*,r_p) \rangle $'s received have  value $0$ 
          in the second component\\
\>\>\> \textbf{then}\\
\>\>\>\> $\send  \langle (P,0,r_p) \rangle $ to all\\
\>\>\> \textbf{else}\\
\>\>\>\>\textbf{if} all the  $\langle (R,*,r_p) \rangle $'s received have value 1 in the second component\\
\>\>\>\>\textbf{then} \\
\>\>\>\>\> $\send  \langle (P,1,r_p) \rangle $ to all\\
\>\>\>\>\textbf{else} \\
\>\>\>\>\> $\send  \langle (P,?,r_p) \rangle $ to all\\
\>\>\> \textbf{wait until} [$\rec \langle (P,*,r_p) \rangle $ from $n-f$ processes] 
                                      (where $*$ can be 0, 1, or ?)\\
\>\>\> \textbf{if} at least  $f\!+\!1$ of the  $\langle (P,*,r_p) \rangle $'s received have the same $w\in\{0,1\}$
         in the second component \\
\>\>\> \textbf{then}\\
\>\>\>\>  $x_p:=w$\\
\>\>\>\>  $\dec(w)$\\
\>\>\> \textbf{else}\\
\>\>\> \>\textbf{if}  one of the $\langle (P,*,r_p) \rangle $'s received have $w\in\{0,1\}$
         in the second component \\ 
\>\>\>\>\textbf{then}\\
\>\>\>\>\> $x_p:=w$\\
\>\>\>\> \textbf{else}\\
\>\>\>\>\> $x_p:=0$\\
\>\>\> $r_p:=r_p +1$\\
\end{tabbing}
\caption{A $^c{C}$-reduction from  \ufT$(\Pi,f)$ to \fT$(\Pi,f)$}
\label{figfnf}
$\hrulefill$
\end{figure*}

\begin{theorem}\label{fnf}
Let $n$ and $ f$  be  two integers such that $1\leq 2f  \leq  n-1$, and let $\Pi$ be a set of $n$
	processes.
The algorithm in Figure~\ref{figfnf} that uses $^c{\cal O}$.\fT$(\Pi,f)$  solves  \ufT$(\Pi,f)$, 
	and so  \ufT$(n,f)$ is $^c{C}$-reducible to \fT$(n,f)$.
\end{theorem}

\begin{Proof}
First notice  that Claims O1 and O2 still hold for any consistent oracle  $^c{\cal O}.$\kT$(\Pi,f)$.

Let $\Pi$ be a set of $n$ processes, and let  $\rho=<\!\! F, I, H \!\!>$ denote a run of the algorithm
	 in Figure~\ref{figfnf} with at most $f$ failures.
By the $f$-resiliency property, the oracle $^c{\cal O}$.\fT$(\Pi,f)$ definitely 
	answers in each of its two consultations in $\rho$.
Let $d$ denote the second answer.
	
To prove that $\rho$ satisfies the termination, irrevocability, agreement, and 
	$(f\!+\!1)$-validity conditions, we shall distinguish the cases $d=0$ 
	and $d=1$.\\

\noindent\textit{Case} $d=1$.  
Irrevocability, termination and agreement are obvious.

Let us  prove that $\rho$ satisfies the $(f\!+\!1)$-validity condition.
\begin{enumerate}
\item Suppose that at least  $f\!+\!1$ processes start with  0.
We consider the first consultation of $^c{\cal O}$.\fT$(n,f)$:
	each of these processes that  starts with 0 either does not query the oracle 
	(because they crash) or queries it with value 0.
By Claim~O1, the first response given by $^c{\cal O}$.\fT$(n,f)$ is necessarily equal to 0, and 
	processes decide 0, as required.
\item Now assume that all the initial values are equal to 1. 
Since $^c{\cal O}$.\fT$(n,f)$ is supposed to be consistent and $d=1$, the oracle may not 
	answer 0 in the first consultation.
Therefore $v=1$ and processes decide 1 in $\rho$.
\end{enumerate}

\noindent\textit{Case} $d=0$.
First, we  prove that $\rho$ satisfies both  the agreement and irrevocability conditions. 
By the  rule which determines when a process proposes value $v\in\{0,1\}$
	(i.e., sends $(P,v,r)$ to all), it is impossible for a process  to propose 0 and for another one to
	propose 1 at the same round.
Suppose now that some processes make a decision in $\rho$;   let $r$ denote the first round
	at which a decision is made, and let $p$ denote a process that decides at round $r$.
Process $p$  has received at least $f\!+\!1$ propositions for its decision value 
	$v$ at round $r$.
Thus every process $q$ receives at least one proposition for $v$ at round $r$,
	and so we have $x_q=v$ at the end of round $r$.
Hence every process that is still alive  decides  $v$ at the latest at round $r+1$, and  keeps
	deciding $v$ in all subsequent rounds.
In other words, $\rho$ satisfies agreement and irrevocability.

For termination, we argue as for the reduction from $\Cons(n,1)$ to $\AC(n,1)$ (see Section~I.8.3).
Since every second query value of $^c{\cal O}$.\fT$(\Pi,f)$ is 1 and $d=0$,  
	by Claim~O2 exactly $f$ failures occur in run $\rho$.
For every process $p$, we consider  the first round  $r_p$  process $p$ executes after the
	last failure occurs in $\rho$, and we let $r_{\rho}=\max_{p\in Correct(F)}(r_p)$.
If some process makes a decision by round $r_{\rho}$, then the above argument for agreement 
	shows that every process that is still alive at  the end of round $r_{\rho}$ has made  a decision by 
	the end of this round.
Suppose no process has made a decision by the end of round $r_{\rho}$.
All correct processes receive the same set of $n-f$ messages of the form $(R,-,r_{\rho})$,
	and so they propose the same value $w\in\{0,1,?\}$ at round $r_{\rho}$.

If $w\neq ?$, then every correct process decides $w$ since it receives $n-f $ propositions
	for $w$ and $n-f\geq f\!+\!1$.
Otherwise, $w=?$ and every correct process $p$ sets its variable $x_p$ to 0 in the end of 
	round $r_{\rho} $.
Since $n-f\geq f\!+\!1$, it is easy to see that every correct process proposes 0  at round $r_{\rho} +1$,
	and so every correct process decides 0.
This completes the proof of termination.

Finally, let us establish  that $\rho$ satisfies the $(f\!+\!1)$-validity condition: 
\begin{enumerate}
\item A simple inductive argument shows that if at least $f\!+\!1$ processes start with value 0, then 
	at any round, value 1 may not be proposed by any process.
Therefore in this case, 0 is the only possible decision value.
\item Now suppose that all processes start with the same initial value $1$.
Every process proposes 1 at the first round, i.e., sends $(P,1,1)$ to all.
As $n-f \geq f\!+\!1$, it follows from the code that each process then decides  $1$.
 \end{enumerate}\qed 
\end{Proof}

As mentioned above,  Theorem~\ref{fnf} states a $^c{C}$-reducibility result which is, in the particular 		case $f=1$, slightly weaker than the $C$-reducibility result given by Theorem~I.8.4.
An open question is  whether  \ufT$(n,f)$ is actually $C$-reducible to \fT$(n,f)$.
If not, this would show in particular that $C$-reduction is strictly stronger than $^c{C}$-reduction. 
	
\section{$\mathbf{C^*}$-reductions between $\mathbf{k}$-Threshold Agreement tasks}

In this section, we establish  two $C^*$-reducibility statements for the $k$-Threshold Agreement 
	tasks which compare \kT$(n,f)$  with \ukT$(n\!+\!1,f)$ and \ukT$(n,f)$ respectively, and contain
	all the $C^*$-reducibility results established in Part~I.
Interestingly, the two $C^*$-reductions that we give here are similar, except in their final 
	decision rule.

\subsection{$\mathbf{C^*}$-reduction when varying  threshold value and number of processes}

Since $\Cons(n,f)$ is  generally not $C$-reducible to $\AC(n,f)$ (cf. Theorem~I.8.2),
	we cannot expect to extend the $C$-reducibility result in Proposition~\ref{generalization}
	from \ukT$(n,f)$ to \kT$(n,f)$ for all the threshold values $k$ less than $f +1$.
Nevertheless, we are going to show that \ukT$(n+1,f)$ is always $C^*$-reducible to \kT$(n,f)$. 

Let $\Pi$ denote the set of $n+1$ processes $\{1,\cdots,n\!+\!1\}$.
We consider the $n\!+\!1$ subsets of $\Pi$ of cardinality $n$, and we denote 
	 $\Pi_{i}=\Pi\setminus \{i\}$.
We use $\overline{\imath}$  as shorthand for $k$-TAg$(\Pi_{i},f)$; 
	hence  $\overline{\imath}$ is the sanctuary of the oracle ${\cal O}$.\kT$(\Pi_{i},f)$ 
	(cf. Section~I.3.3).
	
In Figure~\ref{figc*1}, we give the code of a simple $(k\!+\!1)$-Threshold Agreement  algorithm 
        for $\Pi$ using the oracles ${\cal O}$.\kT$(\Pi_{1},f), \ldots, 
        {\cal O}$.\kT$(\Pi_{n\!+\!1},f)$. 
Informally, every process $i$ consults these oracles with its initial value $v_i$, 
	according  to the order $1,\ldots, n\!+\!1$, except the oracle ${\cal O}$.\kT$(\Pi_{i},f)$
	since $i$ is not a consultant of this oracle ($i\notin \Gamma({\overline{\imath}})$).
As soon as process $i$ gets a response from an oracle, $i$ broadcasts it in $\Pi$.
In this way,  it eventually knows all the values answered by the oracles (including the one by 
	${\cal O}$.\kT$(\Pi_{i},f)$), and then decides on the greatest value.
	
\begin{figure*}[t]\small 
$\hrulefill$
\begin{tabbing}
mm\=mm\=mm\=mm\=mm\=mm\=mm\=mm\=mm\=mm\=mm\= \kill

\textbf{Algorithm for  process} $i:$\\ \\

\textbf{initialization:}\\
\> $d_i\in V\cup \{\bot\}$, initially $\bot$\\ \\
\textbf{for} $l=1$ \textbf{to} $n\!+\!1$ \textbf{do}:\\
\> \textbf{if} $l\neq i$ \textbf{then}\\
\> \> $\query ({\cal O}$.\kT$(\Pi_{l},f)) \langle v_i \rangle $\\
\> \> $\answ ({\cal O}$.\kT$(\Pi_{l},f)) \langle w_l \rangle $\\
\> \> $\send  \langle (l,w_l) \rangle $ to all\\
\textbf{wait until} [$\rec \langle (l, w_l) \rangle $ for all $l\in\{1,\cdots,n\!+\!1\}$]\\
$d_i:=\max _{l=1\cdots,n\!+\!1} (w_l)$\\
$\dec(d_i)$
\end{tabbing}\vspace{-0.5cm}
\caption{A $C^*\!$-reduction from \ukT$(n\!+\!1,f)$ to \kT$(n,f)$.}
\label{figc*1}
$\hrulefill$
\end{figure*}

\begin{proposition}\label{c*1}
If $n, f$ and $k$  are  three integers such that $1\leq f  \leq  n-1$ and $1\leq k  \leq  n$, 
	then the algorithm in Figure~\ref{figc*1} solves the task \ukT$(n\!+\!1,f)$, and so
	\ukT$(n\!+\!1,f)$ is  $C^*$-reducible to \kT$(n,f)$.
\end{proposition}

\begin{Proof}
We first prove the termination property.
By  induction on $i$, we easily show that every oracle ${\cal O}.$\kT$(\Pi_{i},f)$
	is consulted by at least $|\Pi_{i}|-f=n-f$ processes, and so no process is blocked in
	the sanctuary $\overline{\imath}$.
Every correct process $p\in \Pi_{i}$ thus gets an answer from the oracle 
	${\cal O}$.\kT$(\Pi_{i},f)$, and then broadcasts it in $\Pi$.
Since $n\geq f\!+\!1$, each subset $\Pi_{i}$  contains at least one correct process.
Therefore every correct process eventually knows the $n+1$ values answered by the oracles 
	${\cal O}$.\kT$(\Pi_{1},f), \ldots, {\cal O}$.\kT$(\Pi_{n\!+\!1},f)$, 
	and then makes a decision.
	
Irrevocability is obvious.
Agreement follows from the decision rule and the fact that every process which 
	makes a decision knows all the values answered by the oracles.  

For the  validity condition, consider any run of the algorithm with at most $f$ failures.
The proof of termination shows that each oracle  ${\cal O}.$\kT$(\Pi_{i},f)$
	answers to all its consultants.
\begin{enumerate}
\item Suppose that at least $k\!+\!1$ processes in $\Pi$ start with $0$.
In each subset $\Pi_{i}$, at least $k$ processes either do not query the oracle 
	${\cal O}.$\kT$(\Pi_{i},f)$ or query it with value 0.
By Claim~O1,  every oracle necessarily answers 0, and so the decision value is 0.

\item Suppose now that all the all the initial values are 1 and at most $k$ (and so at most $\min( k,f)$)
	processes crash.
Among the subsets  $\Pi_{1},\ldots, \Pi_{n\!+\!1}$ are at least $k$
	with less than $k$ faulty processes.
By Claim~O2,  the corresponding oracles are bound to answer 1.
From the decision rule, it follows that the decision value is 1 since we have $k\geq 1$.
\end{enumerate}
This shows the $(k\!+\!1)$-validity condition.
\qed
\end{Proof}

Note that for $k=n$, Proposition~\ref{c*1} yields 
	$$\Cons(n+1,f)\SC \Cons(n,f),$$ 
	which is also a consequence of the $C$-reduction from $\Cons(n+1,f)$ to $\Cons(n,f)$ that 
	we have shown in Part~I (Proposition~I.7.4).

More interestingly, by Proposition~\ref{c*1} applied $f$ times, we obtain
	$$(f\!+\!1)\mbox{-TAg}(n+f,f)\SC \AC(n,f).$$
Since the task $(f\!+\!1)$-TAg$(n+f,f)$ is equivalent to $\Cons(n+f,f)$ with respect to $\C$, we have 
	$$\Cons(n+f,f) \SC \AC(n,f),$$
	 which is  the first $C^*$-reduction established in Theorem~I.6.4.
	 
Finally, observe that if we were able to strengthen Theorem~\ref{fnf} by proving that \ufT$(n\!+\!f\!-\!1,f)$ 
	is actually $C$-reducible (and not only $^cC$-reducible) to \ufT$(n\!+\!f\!-\!1,f)$ when 
	$n\!+\!f\!-\!1 >2f$, then we could stop  one step before when applying  Proposition~\ref{c*1}.
This would yield the better $C^*$-reducibility result  
$$\Cons(n+f-1,f) \SC \AC(n,f)$$

	when $f\leq n-2$.
	
\subsection{$\mathbf{C^*}$-reduction when  varying the number of processes only}

From Proposition~\ref{c*1}, we cannot derive that $\AC(n\!+\!1,f)$ is $C^*$-reducible to 
	$\AC(n,f)$ (cf. Proposition~I.7.1).
A general statement  for $k$-Threshold Agreement tasks which would extend this latter 
	$C^*$-reducibility result would necessarily compare two tasks with the same threshold value.
	
It turns out that, by just substituting ``$\min$'' for ``$\max$'' in the decision rule 
	in Figure~\ref{figc*1}, the resulting algorithm 
	solves the task \kT$(n\!+\!1,f)$.
This shows that  $k\mbox{-TAg}(n\!+\!1,f)$  is $C^*$-reducible  to $k\mbox{-TAg}(n,f)$.
The proof is similar to the one of Proposition~\ref{c*1}, and is therefore omitted. 

\begin{proposition}\label{c*2}
If $n, f$ and $k$  are  three integers such that $1\leq f  \leq  n-1$ and $1\leq k  \leq  n$, 
	then the task \kT$(n\!+\!1,f)$ is $C^*$-reducible to \kT$(n,f)$.
\end{proposition}

Specializing $k$ to 1 in Proposition~\ref{c*2}, we actually recover that  $\AC(n\!+\!1,f)$ is 
	$C^*$-reducible to $\AC(n,f)$.
Note that when $k\geq f\!+\!1$, Proposition~\ref{c*1} can be derived from 							Propositions~\ref{generalization} and~\ref{c*2}.

\section{C-irreducibility results between k-Threshold Agreement tasks}

In this section, we shall examine generalizations of the two $C$-irreducibility results between 
	Consensus and Atomic Commitment tasks established in Part~I.
More precisely, we shall prove that for a fixed set $\Pi$ of $n$ 	processes, and a fixed resiliency
	degree $f$, $1\leq f \leq n\!-\!1$, the task  \kT$(\Pi,f)$  is incomparable with
	$\Cons(\Pi,f)$ with respect to $\C$ for any threshold value $k\in\{1,\ldots,f\!-\!1\}$.
	
By Propositions~\ref{generalization} and~\ref{eqcons}, we know that the   tasks
	 \ufT$(\Pi,f),\ldots , n\mbox{-TAg}(\Pi,f)=\Cons(\Pi,f)$ are all of  the same unsolvability 
	 degree. 
Thus  it remains to compare the task $f\mbox{-TAg}(\Pi,f)$ with $\Cons(\Pi,f)$ to get a complete 
	picture of the relationships between the various $k$-Threshold Agreement tasks for a fixed 
	set of processes and a fixed resiliency  degree.
	
To some  extent, Theorem~\ref{fnf} answers this question when a majority of processes 
	is correct ($n>2f$), since it establishes that $\Cons(\Pi,f)$ is $^cC$-reducible to  \fT$(\Pi,f)$.
We shall prove that this does not hold anymore with a minority of correct processes:  if $n\leq 2f$, 
	then  $\Cons(\Pi,f)$ is not  $^cC$-reducible (and so not $C$-reducible) to  \fT$(\Pi,f)$.
	
Besides providing a better understanding of the connections between the various  tasks
	$k\mbox{-TAg}(\Pi,f)$,
	the three irreducibility results that we  establish in this section  will play a key role in
	the proofs of our final results comparing wait-free and $f$-resilient 
	Consensus tasks (see Section~\ref{waitfres} {\it infra}).	 

\subsection{C-irreducibility to wait-free Consensus tasks}

\begin{theorem}\label{ir1}
For every integers $n,f,k$  such that $1\leq k\leq f \leq n-1$,  the task \kT$(n,f)$ 
	 is not  $C$-reducible to $\Cons(n,n-1)$, and so is not  $C$-reducible to $\Cons(n,f)$.
\end{theorem}

\begin{Proof}
Let $\Pi$ be a set of $n$ process names. 
Suppose, for the sake of contradiction, that there is an algorithm $R$ for the task \kT$(\Pi,f)$ 
	which uses the oracle ${\cal O}.\Cons(\Pi,n-1)$. 
Let $\Pi_1$ be any subset of $\Pi$ of cardinality $k$.
Consider a run $\rho=<\!\! F, I, H \!\!>$ of $R$ such that, for any $q\in\Pi$, 
	$I(q)=s_q^1$, and for any $t\in {\cal T}$, $F(t)=\Pi_1$. 
In other words, $\rho$ is a run of $R$ in which all processes start with
	initial value 1 and no process is faulty except the processes in $\Pi_1$, all of  which 
	initially crash.
Since $k\leq f$, every process in $\Pi\setminus\Pi_1$ eventually  makes a decision in $\rho$, and all the 
	decision values are identical; let $d$ denote this common decision value. 

We now introduce  the mapping $I'$ which is identical to $I$ over $\Pi\setminus \Pi_1$ and
	satisfies $I'(p)=s_p^0$ for any process $p$ in $\Pi_1$.
Then we consider $\rho'=<\!\! F, I', H \!\!>$;  we claim that $\rho'$ is a run of $R$.
Recall that the runs of $R$ are defined by the compatibility rules R1--6 introduced in 
	Section~I.4.2.
	
Since $\rho$ is a run of $R$, it is straightforward that $\rho'$ satisfies R1, R2, R3, R5, and R6.
By an easy induction, we see that for any process $q$, $q\neq p$, the sequence 
	of the local states reached by $q$ are the same in $\rho'$ as in $\rho$.
This ensures that every step in $H$ is feasible from $I'$, and so R4 holds in $\rho'$. 
Thus, $\rho'$ is a run of $R$, and by the $k$-validity condition, 
	the only possible decision value in $\rho'$ is 0. 
This shows that $d=0$. 

Now from $\rho$, we are going to construct a failure free run of $R$ by using the asynchronous 
	structure of computations.
To achieve that, we need the following lemma, where  $F_0$ denotes  the failure pattern 
	with no failure (defined formally by $F_0(t)=\emptyset$, for any $t\in {\cal T}$), and
	$H[0,t]$ denotes the prefix in $H$ of events with time less or equal to $t$.
		
\begin{lemma}\label{ext}
For any $t_0\in {\cal T}$, there exists an extension $H_0$ of $H[0,t_0]$
	such that $<\!\! F_0, I, H_0 \!\!>$ is a failure free run of $R$.
\end{lemma}
\begin{Prooflemma} 
The history $H_0$ is constructed in stages, starting from $H[0,t_0]$;
	each stage consists in adding  one event.
A queue of the processes in $\Pi$ is maintained, initially in an arbitrary order, and
	the messages in $\beta$ are  ordered according to the time the messages were
	sent, earliest first.
	
Suppose that the finite history $H_0[0,t]$ extending $H[0,t_0]$ is constructed.
Let $t^+$ denote the successor of $t$ in ${\cal T}$, and let $q$ be the first process 
	in the process queue.
After $H_0[0,t]$, $q$ may achieve only one type $T$ of event. 
There are three cases to consider:

\begin{enumerate}
\item $T=S$ or $T=Q$.
The automaton $R(q)$ entirely determines the event $e=(\beta, q,t^+,S,m)$ or 
	$e=(\Cons(\Pi,n-1), q,t^+,Q,v)$ which $q$ may achieve at time $t^+$.
\item $T=R$.
In this case, the message buffer $\beta$ contains at least one message for $q$.
Then we let $e= (\beta, q,t^+,R,m)$, where $m$ denotes the earliest message for $q$ 
	in $\beta$.
\item $T=A$.
Form the successive consultations of ${\cal O}.\Cons(\Pi,n-1)$ in $H_0[0,t]$, and focus on 
	 the latter consultation. 
Note that  process $q$ has necessarily queried ${\cal O}.\Cons(\Pi,n-1)$
	during  this consultation; let $v$ be the value of this query.
There are three subcases:
	\begin{description}
	\item {\em Case 1:}  ${\cal O}.\Cons(\Pi,n-1)$  has already answered some value $d$.\\
	In this case, we let $e=(\Cons(\Pi,n-1), q,t^+,A,d)$.
	\item {\em Case 2:}  ${\cal O}.\Cons(\Pi,n-1)$ has not answered yet.\\
	We let $e=(\Cons(\Pi,n-1), q,t^+,A,v)$.
	\end{description}
\end{enumerate}

The above procedure determines a unique event $e$, and we let $H_0[0,t^+] = H_0[0,t]; e$ (where 
	semicolon denotes concatenation). 
Process $q$ is then moved to the back of the process queue.

This  inductively defines $H_0$.
By construction,   $\rho_0=<\!\! F_0, I, H_0 \!\!>$ satisfies R1--6, and so is a failure free run 
	of $R$. \nopagebreak \hfill $\Box_{Lemma~\ref{ext}}$ \nopagebreak
\end{Prooflemma}

We now instantiate  $t_0$  to be the time when the last process makes a decision in $\rho$.
The lemma provides an extension $H_0$ of $H[0,t_0]$
	such that $\rho_0=<F_0,I,H_0>$ is a run of $R$.
The decision value in $\rho_0$ is 0, which contradicts the fact that processes 
	must decide on 1 in any  failure free run of an algorithm solving \kT$(\Pi,f)$ in 
	which all processes start with initial value 1.\qed
\end{Proof}

In the case $f=1$, Theorem~\ref{ir1} states that $\AC(n,1)$ is not $C$-reducible to
	$\Cons(n,n-1)$,  and so reduces to Theorem I.8.1.

\subsection{C-irreducibility to wait-free k-TAg tasks}

\begin{theorem}\label{ir2}
For every integers $n,k$ such that $2\leq k\leq n-1$, $\Cons(n,k)$
	is not $C$-reducible to \lT$(n,n-1)$. 
\end{theorem}

\begin{Proof}
We  proceed by contradiction: let $\Pi$ be a set of $n$ process names, and suppose that 
	there is an algorithm $R$ for $\Cons(\Pi,k)$ using  the oracle ${\cal O}$.\lT$(\Pi,n\!-\!1)$.
Recall that the sanctuary of this oracle is \lT$(\Pi,n\!-\!1)$ itself  (cf. Section~I.3.3);
	to simplify notation, we let $\sigma=$\lT$(\Pi,n\!-\!1)$.

We fix some subset $\Pi_1\subseteq \Pi$ of cardinality $k-1$, and we denote 
	$\Pi'=\Pi\setminus \Pi_1$.
From $R$, we shall design an algorithm $A$ running on $\Pi'$, which uses no oracle.
Then we  shall prove that $A$ solves the task $\Cons(\Pi',1)$, which contradicts the 
	impossibility of Consensus with one failure established by Fischer, Lynch, and 
	Paterson~\cite{FLP85} since $|\Pi'| =n-(k-1)\geq 2$.  
	
For each process $q$ in $\Pi'$, we define the automata $A(q)$ in the following way:  	
	\begin{itemize}
	\item the set of states of $A(q)$ is the same as the one of $R(q)$;
	\item the set of initial states of $A(q)$ is the same as the one of $R(q)$;
	\item each transition $(s_q,[q,m,\bot], s'_q)$ of $R(q)$ in which $q$ consults no oracle
	 is also a transition of $A(q)$;
	\item each transition $(s_q,[q,m,1], s'_q)$ of $R(q)$ in which the oracle 
	answers 1 is removed;
	\item each transition $(s_q,[q,m,0], s'_q)$ of $R(q)$ in which the oracle 
	answers 0 is replaced by the transition $(s_q,[q,m,\bot], s'_q)$.
	\end{itemize}
Note that all the steps in  $A(q)$ are of the form  $[q,m,\bot]$; in other words, the algorithm 
	$A$ uses no oracle.

Let $\rho_A=<\!\! F, I, H \!\!>$ be any run of $A$.
Since $A$ uses no oracle, each event  in $H$  is of the form $e=(\beta,q,-,-,-)$ 
	and is part of  some  transition 
	$(s_q, [q,m,\bot] ,s'_q)$ of $A(q)$, where $m\in M\cup \{ \mbox{null} \}$.
In the construction of $A(q)$ described above, this transition results from some unique transition 
	of $R(q)$, of the form $(s_q, [q,m,\bot] ,s'_q)$ or $(s_q, [q,m,0],s'_q)$.
In this way, to each event in $H$, we associate a unique transition of $R(q)$ in which the oracle
	at sanctuary $\sigma$ is not consulted or answers 0.
	
Now, to each run $\rho_A=<\!\! F, I, H \!\!>$ of $A$, we associate the triple  
	$\rho_R=<\!\! F', I', H' \!\!>$, where the failure pattern $F'$ is defined by 
	$$F' :  t\in {\cal T}\rightarrow F'(t)=F(t)\cup\Pi_1,$$
	the mapping $I'$ by: 
	\begin{enumerate}
	\item for any process $q\in \Pi'$, $I'(q)=I(q)$;
	\item for any $q\in \Pi_1$, we let $I'(q)=s^0_q$ if  $I(p)=s^0_p$ for some process $p\in \Pi'$; 
	otherwise we let $I'(q)=s^1_q$,
	\end{enumerate}
and the sequence $H'$ is constructed from $H$ by the following rules:
	\begin{enumerate}
	\item any event in $H$ that is associated to a transition of $R$ in which 
	the oracle is not consulted is left unchanged;
	\item any event $(\beta,q,t,\mbox{R},m)$ in $H$, even when associated to some transition 
	in $R(q)$ in which $\sigma$ is consulted, is  left unchanged; 
	\item an event $(\beta,q,t,\mbox{S},m)$ in $H$ which is associated to some 
	transition in $R(q)$ of the form  $(s_q, [q,-,0], s'_q)$, is replaced  in $H'$ by 
	the three events series 
	$$\langle (\sigma,q,t,\mbox{Q},v),(\sigma,q,t,\mbox{A},0),(\beta,q,t,\mbox{S},m)\rangle,$$
	where $v$ is the query value determined by  $s_q$.
	 \end{enumerate}

We claim that the triple $\rho_R$ so defined is a run of $R$.
By construction of $H'$, there is no event in $H'$ whose process name is in $\Pi_1$,
	and each event in $H'$ at time $t$ corresponds to at least one event in $H$
	that also occurs at time $t$.
Since $H$ is compatible with $F$ and $F'(t)=F(t)\cup \Pi_1$, it follows that $H'$
	is compatible with $F'$.
For any process $q\in \Pi'$,  $H|q$ is well-formed, and so  is $H'|q$.
This proves that $H'$ satisfies R2.

From the R3, R4, and R6 conditions for $H$, it is also immediate to prove that in turn
	$H'$ satisfies  R3, R4, and R6.
	
Now since $F(t)\subseteq F'(t)$, every process $q$ which is correct in $F'$ is also correct
	in $F$, and so takes an infinite number of steps in $H$.
By construction of $H'$, it follows that $q$ takes an infinite number of steps in $H'$.
Thus $H'$ satisfies R5.

Finally, to show that $\rho_R$ satisfies R1, we  focus on a consultation of $\sigma$ in $H'$. 
By construction of $H'$, the only  value answered by the oracle at sanctuary $\sigma$  is 0.
This trivially enforces agreement.
Since there are at least  $k-1$  faulty processes in $F'$, 
	the answer 0 is allowed for $F'$ and any input vector $\vec{V}\in \{0,1\}^{\Pi}$
	with regard to the $(k\!-\!1)$-validity condition.
Besides, every step in $H$ is complete (with a receipt and a state change), and so 
	by construction of $H'$, the oracle answers in each consultation of $H'$. 
It follows that $H'|\sigma$ is an history of the oracle ${\cal O}$.\lT$(\Pi,n\!-\!1)$.
This completes the proof that $\rho_R=<\!\! F', I', H' \!\!>$  is a run of $R$.

Let $\rho_A$ be any run of $A$ with at most one failure;
	 in the corresponding run $\rho_R$ of $R$,  at most $1+(k-1)=k$ processes fail .
Since $R$ is an algorithm that solves  $\Cons(\Pi,k)$,
	$\rho_R$ satisfies the termination, agreement, irrevocability and validity 
	conditions of Consensus.
It immediately follows that the run $\rho_A$, which $\rho_R$ stems from, also satisfies 
	the termination, agreement, and irrevocability conditions.
Moreover, by definition of $I'$, if all processes start with the same initial value 
	$v$ in $\rho_A$, then they also have the same initial value $v$ in $\rho_R$; 
	the only possible decision value in $\rho_R$, and so in $\rho_A$, is $v$.
	
Consequently,  $A$ is an algorithm for $\Cons(\Pi',1)$  using no oracle, a contradiction 
 	with~\cite{FLP85}.\qed
 \end{Proof}

Notice that for $k=2$, Theorem~\ref{ir2} states that $\Cons(n,2)$ is not $C$-reducible to $\AC(n,n-1)$,
	and so reduces to Theorem I.8.2.
	
Importantly, we may safely  substitute the consistent oracle $^c{\cal O}$.\lT$(\Pi,n\!-\!1)$
	for  ${\cal O}$.\lT$(\Pi,n\!-\!1)$ in the proof of Theorem~\ref{ir2}.
In this way,  we prove a result slightly stronger than Theorem~\ref{ir2} by establishing that 
	$\Cons(n,k)$ is actually not $^cC$-reducible to \lT$(n,n-1)$.

\begin{corollary}\label{corir2}
For any integers $k, f, n$ such that $1\leq k\leq f\!-\!1$ and $f\leq n\!-\!1$, $\Cons(n,f)$ is not $^cC$-reducible
	to \kT$(n,f)$.
\end{corollary} 

\begin{Proof}
Suppose, for the sake of contradiction, that $\Cons(n,f)$ is $^cC$-reducible to some \kT$(n,f)$
	with $1\leq k \leq f-1$.
Since $f\geq k\!+\!1$,  $\Cons(n,k\!+\!1)$ is a special case of  $\Cons(n,f)$, and so 
	$$\Cons(n,k\!+\!1) \leq_{^cC} \Cons(n,f).$$
Similarly,  \kT$(n,f)$ is a special case of \kT$(n,n\!-\!1)$, and we have
	$$k\mbox{-TAg}(n,f) \leq_{^cC} k\mbox{-TAg}(n,n\!-\!1).$$
Using  transitivity of the $^cC$-reduction, we obtain  that $\Cons(n,k\!+\!1)$ is $^cC$-reducible to 
	$k\mbox{-TAg}(n,n\!-\!1)$, a contradiction with the variant of Theorem~\ref{ir2} alluded above.\qed
\end{Proof}

\subsection{A C-irreducibility  result when a majority of processes may fail}

We now complete the comparison between the various $k$-Threshold Agreement tasks 
	for a fixed set of processes $\Pi$ and a fixed resiliency degree $f$.
We are going to prove that if a majority of processes may be faulty ($|\Pi|\leq 2f$), then 
	$\Cons(\Pi,f)$ is not $C$-reducible to \fT$(\Pi,f)$.
Combining this latter irreducibility result with Theorem~\ref{ir1}, we conclude that  with respect to $\C$,
	\fT$(\Pi,f)$ is incomparable with
 	any of the equivalent tasks 
	\ufT$(\Pi,f),\ldots,n$-TAg$(\Pi,f)=\Cons(\Pi,f)$. 
	
\begin{theorem}\label{ir3}
Let $n$ and $f$ be two integers such that $1\leq f\leq n-1$.
If $n\leq 2f$, then the task \ufT$(n,f)$ is not $C$-reducible to \fT$(n,f)$,
	and so $\Cons(n,f)$ is not $C$-reducible to \fT$(n,f)$.
\end{theorem}

\begin{Proof}
Let $\Pi$ be any set of $n$ processes and let $f$ denote an integer such that 
	$1\leq f \leq n-1$ and $n\leq 2f$.
Suppose, for the sake of contradiction that there is an algorithm $R$ which solves 
	\ufT$(\Pi,f)$ using the oracle ${\cal O}.$\fT$(\Pi,f)$.
	
We partition $\Pi$ into two sets $\Pi'$ and $\Pi''$ such that $\Pi'$ contains $f$ processes,
	and $\Pi''$ contains the remaining $n-f$ processes.
Since $n>f$, we have $f>2f-n$; 
	we fix any (possibly empty) strict subset $\pi'$ of $\Pi'$ with  $2f-n$ processes.

Consider the triple $\rho'=<\!\! F', I, H' \!\!>$  where the failure pattern $F'$ is defined by 
	$$F' :  t\in {\cal T}\rightarrow F'(t)=\pi'\cup \Pi'',$$
	the mapping $I$ by: 
	\begin{enumerate}
	\item for any  process $p\in \Pi'$, $I(p)=s^0_p$
	\item for any process $p\in \Pi''$, $I(p)=s^1_p$;	
	\end{enumerate}
and the sequence $H'$ is constructed by induction on $t\in {\cal T}$, as follows.

First, $H'[0,0]$ is defined to be the empty sequence.
A queue of the processes in $\Pi'\setminus\pi'$ is maintained, initially in an arbitrary order, and
	the messages in $\beta$ are  ordered according to the times the messages were
	sent, earliest first.
Suppose that the finite history $H'[0,t]$  is constructed.
Let $t^+$ denote the successor of $t$ in ${\cal T}$, and let $q$ be the first process 
	in the process queue.
After $H'[0,t]$, $q$ may execute only one type $T$ of event. 
There are three cases to consider:

\begin{enumerate}
\item $T=S$ or $T=Q$.
The automaton $R(q)$ entirely determines the event $e=(\beta, q,t^+,S,m)$ or 
	$e=(f\mbox{-TAg}(\Pi,f), q,t^+,Q,v)$ that $q$ may execute  at time $t^+$.
\item $T=R$.
In this case, the message buffer $\beta$ contains at least one message for $q$.
Then we let $e= (\beta, q,t^+,R,m)$, where $m$ denotes the earliest message for $q$ 
	in $\beta$.
\item $T=A$.
	We let $e=(f\mbox{-TAg}(\Pi,f), q,t^+,A,0)$.
\end{enumerate}

The above procedure determines a unique event $e$, and we let $H'[0,t^+] = H'[0,t]; e$.
Process $q$ is then moved to the back of the process queue.
This  inductively defines $H'$.

\begin{lemma}\label{rho'}
The triple $\rho'=<\!\! F', I , H' \!\!>$ is a run of $R$ in which every process in 
	$\Pi'\setminus\pi'$ decides 0.
\end{lemma}

\begin{Prooflemma} 
By the definitions of $F'$ and $H'\!$, it is immediate that $\rho'$ satisfies properties R2-6.
For R1, the only non-trivial point is checking that $H'|$\fT$(\Pi,f)$ satisfies the $f$-validity 
	condition, or in other words 
	that  the oracle is always allowed to answer 0.
For that, we just need to observe that any process in $\pi'\cup\Pi''$ takes no step in $H'$, 
	and never queries the oracle ${\cal O}$.\fT$(\Pi,f)$.
Hence,   $(n\!-\!f)\!+\!(2f\!-\!n)=f$ processes do not query  the oracle, which is thus allowed to answer 0 
	with respect to the $f$-TAg specification.
It follows that $\rho'$ satisfies R1, and so is a run of $R$.

Because $R$ solves the task \ufT$(\Pi,f)$ and the number of faulty processes in $\rho'$ is $f$,
	all the processes in $\Pi'\setminus\pi'$ make the same decision  in $\rho'$.
Let $d'$ denote the common decision value in $\rho'$.

Consider the mapping $I_0$ such that for any process $p$ in $\Pi$, $I_0(p)=s^0_p$.
Since every process in $\Pi''$ initially crashes in the failure pattern $F'$ and 
	$\rho'$ is a run of $R$ with the decision value $d'$,  the triple 
	$<\!\! F', I_0, H' \!\!>$ is also a  run of $R$ in which $f$ processes are faulty
	and the decision value is $d'\!$.
By the $(f\!+\!1)$-validity condition, the  decision value in this second run of $R$ is equal to 0.
Thus we derive that $d'=0$. \nopagebreak \hfill $\Box_{Lemma~\ref{rho'}}$ \nopagebreak
\end{Prooflemma}

Let $\theta'$ denote the first time when all processes in $\Pi'\setminus\pi'$ have made 
	a decision in $\rho'$. 
Now, consider the triple $\rho''=<\!\! F'', I, H'' \!\!>$  where the failure pattern $F''$ is defined by 
	$$F'' :  t\in {\cal T}\rightarrow F''(t)= \Pi',$$
	and the sequence $H''$ is constructed in the same way as $H'$ with the additional requirement
	that the time of each event in $H''$ is greater than $\theta'$.
A proof similar to the one of Lemma~\ref{rho'} shows the following:

\begin{lemma}\label{rho''}
The triple $\rho''=<\!\! F'', I, H'' \!\!>$ is a run of $R$ in which every process in $\Pi''$ decides~1.
\end{lemma}

Let $\theta''$ denote the first time when all the processes in $\Pi''$ have made a decision in $\rho''$. 
For any $t\in{\cal T}$, we let 
$$\begin{array}{llll}
F(t) & = & \emptyset & \mbox{ when } 0\leq t\leq \theta''\\
         & = & \Pi' & \mbox{ when }  t > \theta''.
\end{array}$$
By construction, the time of each event in $H''$ is greater than $\theta'$, and so 
	we may form the finite history 
	$H'[0,\theta'] ; H''[0,\theta'']$. 

\begin{lemma}\label{ext2}
There exists an extension $H$ of $H'[0,\theta'] ; H''[0,\theta'']$ such that $<\!\! F, I, H \!\!>$ is a  run of $R$.
\end{lemma}

\begin{Prooflemma}
The proof technique is similar to the one of Lemma~\ref{ext}.
The history $H$ is constructed in stages, starting from $H'[0,\theta'] ; H''[0,\theta'']$;
	each stage consists in adding  one event.
A queue of the processes in $\Pi''$ is maintained, initially in an arbitrary order, and
	the messages in $\beta$ are  ordered according to the times the messages were
	sent, earliest first.
	
Suppose that the finite history $H[0,t]$ extending $H'[0,\theta'] ; H''[0,\theta'']$ is constructed.
Let $t^+$ denote the successor of $t$ in ${\cal T}$, and let $q$ be the first process 
	in the process queue.
After $H[0,t]$, process $q$ may achieve only one type $T$ of event. 
There are three cases to consider:

\begin{enumerate}
\item $T=S$ or $T=Q$.
The automaton $R(q)$ entirely determines the event $e=(\beta, q,t^+,S,m)$ or 
	$e=($\fT$(\Pi,f), q,t^+,Q,v)$ which $q$ may achieve at time $t^+$.
\item $T=R$.
In this case, the message buffer $\beta$ contains at least one message for $q$.
Then we let $e= (\beta, q,t^+,R,m)$, where $m$ denotes the earliest message for $q$ 
	in $\beta$.
\item $T=A$.
In this latter case, we let $e=($\fT$(\Pi,f), q,t^+,A,0)$. 
\end{enumerate}

The above procedure determines a unique event $e$, and we let $H[0,t^+] = H_0[0,t]; e$. 
Process $q$ is then moved to the back of the process queue.

This  inductively defines $H$.
By construction,   the triple $\rho=<\!\! F, I, H \!\!>$ satisfies R2--6.
Because every process in $\Pi'$ is faulty and $|\Pi'|=f$, the oracle ${\cal O}$.\fT$(\Pi,f)$ is
	always allowed to answer 0 whatever query values are.
This ensures that $\rho$ satisfies R1.
Consequently $\rho$ is a  run of $R$.\nopagebreak \hfill $\Box_{Lemma~\ref{ext2}}$ \nopagebreak 
\end{Prooflemma}

So, we have just shown that any algorithm using ${\cal O}$.\fT$(\Pi,f)$ for \ufT$(\Pi,f)$ would 
	have a run in which  processes in $\Pi'\setminus \pi'$ decide 0 and processes in $\Pi''$
	decide 1.
Since $f\leq n-1$, both $\Pi'\setminus \pi'$ and $\Pi''$ are non-empty;
	we then conclude that this run  violates the agreement property, 
	a contradiction.\qed
\end{Proof}

Observe that the latter proof crucially relies on the fact
	 that the oracle ${\cal O}$.\fT$(\Pi,f)$
	is allowed to answer on the grounds of informations concerning future failures:
	whatever the query values are, ${\cal O}$.\fT$(\Pi,f)$ may answer 0 
	from the beginning in history $H$ whereas no process
	will crash  before time $\theta''$.
In the next section, we shall actually prove that this $C$-irreducibility result does not hold 
	anymore when considering oracles that do not see into the future.

\subsection{Sham oracles}

We now describe an algorithm that solves \ufT$(\Pi,f)$ with the help of some $f$-resilient
	 oracle suitable for the agreement problem $f$-TAg for $\Pi$ which does not see into the future.
This reduction algorithm works for any resiliency degree, even when a majority of processes
	may fail, and so this mitigates the irreducibility result in Theorem~\ref{ir3} above.
	
First, we formally define such oracles.
Let $\Pi$ be a set of processes, and let $F$ be any failure pattern for $\Pi$.
For any $\theta\in {\cal T}$, $F_{\theta}$ denotes the failure pattern for $\Pi$ defined by
$$F_{\theta}\ :\ t\in{\cal T} \rightarrow \left\{ \begin{array}{ll}
                                                                            F(t) & \mbox{if $0\leq t \leq \theta$}\\
                                                                            F(\theta) & \mbox{otherwise.}
                                                                            \end{array} \right.$$                                                                         
\begin{definition}
Let  ${\cal O}_{\sigma}$ be an oracle whose set of consultants is $\Pi$.
We say that ${\cal O}_{\sigma}$ is  a {\em sham oracle} if for any failure pattern $F$ 
	for $\Pi$, any history $H \in{\cal O}_{\sigma}(F)$, and  any time $\theta\in {\cal T}$, 	
	there exists an extension $H'$ of $H[0,\theta]$ such that $H'\in{\cal O}_{\sigma}(F_{\theta})$.
\end{definition}

As for consistent oracles, for every task $T=(P,f)$, we define the sham version of the oracle 
	for $T$, denoted $^s{\cal O}.T$,  as
	the most general sham oracle which is $f$-resilient and suitable for $P$.
For any failure pattern $F$ for $\Pi$, we have $^s{\cal O}.T(F)\subseteq {\cal O}.T(F)$.
In other words,   $^s{\cal O}.T$ responses  with less scope 
	than  ${\cal O}.T$, and thus the answers given by  $^s{\cal O}.T$ may be thought 
	as  more precise than the ones given by  ${\cal O}.T$.
So, the sham oracle $^s{\cal O}.T$ is ``at least as powerful as''  ${\cal O}.T$ in the sense that 
	any algorithm using ${\cal O}.T_2$ for some task $T_1$ still solves $T_1$ when using
	$^s{\cal O}.T_2$.

This leads to a new notion of reduction {\em \`a la Cook}, called ${^sC}$-reduction and denoted
	$\leq_{^sC}$,  
	in which algorithms may only use sham oracles.
Formally,  $T_1\leq_{^sC} T_2$ if there is an algorithm for $T_1$ using the 
	sham oracle $^s{\cal O}.T_2$.
Since $^s{\cal O}.T$ is at least as powerful as ${\cal O}.T$, $C$-reducibility implies
	$^sC$-reducibility.	

\begin{figure*}[t]\small 
$\hrulefill$
\begin{tabbing}
mm\=mm\=mm\=mm\=mm\=mm\=mm\=mm\=mm\=mm\=mm\= \kill
\textbf{Algorithm for  process} $p:$\\ \\
\> $\query (^s{\cal O}.$\fT$(\Pi,f)) \langle v_p \rangle $\\
\> $\answ (^s{\cal O}.$\fT$(\Pi,f)) \langle v \rangle $\\
\> $\query(^s{\cal O}$.\fT$(\Pi,f)) \langle 1  \rangle $\\
\> $\answ(^s{\cal O}$.\fT$(\Pi,f)) \langle d \rangle $\\
\>\textbf{if} $d=1$\\
\>\textbf{then}\\
\>\> $\dec(v)$\\
\>\textbf{else}\\
\>\> $\send  \langle v_p \rangle $ to all\\
\>\> \textbf{wait until} [$\rec \langle * \rangle $ from $n-f$ processes] 
                                      (where $*$ can be 0 or 1)\\
\>\> \textbf{if} at least  one of the received  values is $0$ \\
\>\> \textbf{then}\\
\>\>\> $\dec(0)$ \\
\>\> \textbf{else}\\
\>\>\> $\dec(1)$ 
\end{tabbing}
\caption{A $^s{C}$-reduction from  \ufT$(\Pi,f)$ to \fT$(\Pi,f)$}
\label{figsored}
$\hrulefill$
\end{figure*}

\begin{theorem}\label{fnfsored}
Let $n$ and $ f$  be  two integers such that $1\leq f  \leq  n-1$, and let $\Pi$ be a set of $n$
	processes.
The algorithm in Figure~\ref{figsored} that uses $^s{\cal O}$.\fT$(\Pi,f)$  solves  \ufT$(\Pi,f)$, 
	and so  \ufT$(n,f)$ is $^s{C}$-reducible to \fT$(n,f)$.
\end{theorem}

\begin{Proof}
First notice  that Claims O1 and O2 also hold for any sham oracle  $^s{\cal O}.$\kT$(\Pi,f)$.

Let $\Pi$ be a set of $n$ processes, and let  $\rho=<\!\! F, I, H \!\!>$ denote a run of the algorithm
	 in Figure~\ref{figsored} with at most $f$ failures.
By the $f$-resiliency property, the oracle $^s{\cal O}$.\fT$(\Pi,f)$ definitely 
	answers in each of its two consultations in $\rho$.
Let $\theta_1$ and $\theta_2$ denote the first time when $^s{\cal O}$.\fT$(\Pi,f)$
	answers in the first and second consultation, respectively.

Termination and  irrevocability are obvious.	
To prove that $\rho$ satisfies the agreement and 
	$(f\!+\!1)$-validity conditions, we shall distinguish the cases $d=0$ 
	and $d=1$ as we did for the $^cC$-reduction in Theorem~\ref{fnf}.\\

\noindent\textit{Case} $d=1$.  
In this case, it is immediate that $\rho$ satisfies the  agreement condition.
Now we prove that $\rho$ satisfies the $(f\!+\!1)$-validity condition.
\begin{enumerate}
\item Suppose that at least  $f\!+\!1$ processes start with  0.
Concerning the first consultation of $^s{\cal O}$.\fT$(\Pi,f)$,
	each of these processes which starts with 0 either does not query the oracle 
	(because it crashes) or queries it with value 0.
By Claim~O1, the first response given by $^s{\cal O}$.\fT$(\Pi,f)$ is necessarily equal to 0, and 
	processes decide 0, as required.
\item Now assume that all the initial values are equal to 1. 
By Claim~O1, at most $f-1$ processes do not query $^s{\cal O}$.\fT$(\Pi,f)$ twice,
	since the second answer $d$ given by  the oracle  is 1.
From the definition of a well-formed oracle history, it follows that no process 
	queries  $^s{\cal O}$.\fT$(\Pi,f)$ for the second time before time $\theta_1$.
Thus, at most $f-1$ processes crash by time $\theta_1$ in $F$, and so 
	$$|Faulty(F_{{\theta}_1})|\leq f-1.$$
Because  $^s{\cal O}$.\fT$(\Pi,f)$ is a sham oracle, it is not allowed to answer 0 in the
	first consultation.
Therefore, $v=1$ and processes decide 1 in $\rho$.
\end{enumerate}

\noindent\textit{Case} $d=0\,$.
Since every query value is 1 in the second consultation  of $^s{\cal O}$.\fT$(\Pi,f)$,
	and  $^s{\cal O}$.\fT$(\Pi,f)$ is a sham oracle,  we do know that in this case, at least 
	$f$ processes have crashed by time $\theta_2$.
At most $f$ processes are faulty in $F\!$, and so exactly $f$ processes have crashed by time 
	$\theta_2$.
Thus, every correct process receives the same set of $n-f$ initial values.
This ensures agreement.
We easily check that $\rho$ satisfies the $(f\!+\!1)$-validity condition.\qed 
\end{Proof}

Combining Theorems~\ref{ir3} and~\ref{fnfsored}, we derive that among agreement tasks, the 
	$C$-reduction defines a strictly  finer hierarchy  than the $^sC$-reduction.
In other words,  the sham oracle $^s{\cal O}.T$ is in general more 
	powerful than ${\cal O}.T$: to be unable to see into the future actually helps to make a decision!
	
Note that the two oracles  $^c{\cal O}.T$ and  $^s{\cal O}.T$ are generally not comparable.
However, we easily check that the reduction from \ufT$(\Pi,f)$ to \fT$(\Pi,f)$ described 
	in Section~\ref{consred} still works when the reduction algorithm uses $^s{\cal O}$.\fT$(\Pi,f)$
	instead of $^c{\cal O}$.\fT$(\Pi,f)$: the fact that $^s{\cal O}$.\fT$(\Pi,f)$ does not see into the future
	makes it sufficiently consistent and ensures that it cannot answer 0 and then 1 with all the
	query values equal to~1.
Conversely, note that the reduction described above does not work when substituting 
	$^c{\cal O}$.\fT$(\Pi,f)$ for $^s{\cal O}$.\fT$(\Pi,f)$.	

\section{Wait-freedom vs. f-resiliency for Consensus tasks}\label{waitfres}

In the previous section, we have established various irreducibility results between pairs of
	$k$-Threshold Agreement tasks,  only one of which is a Consensus task.
Relying on these results, we are now in position to establish $C$-irreducibility results between 			Consensus tasks only.
More precisely, we shall derive two $C$-irreducibility results  between wait-free and $f$-resilient 
	Consensus tasks first for a fixed set of processes, and then for a fixed resiliency degree.
In both cases, we shall show that with respect to $C$-reduction, wait-free Consensus is strictly 
	harder to solve than (non wait-free) $f$-resilient Consensus.
We shall discuss the relationship between our results and previous ones established in the 
	message passing model (\cite{CT96, CHT96}) and in the shared objects model 
	(\cite{BG93, BGLR01, CHJT94,CHJT04}).

\subsection{Wait-freedom and f-resiliency for a fixed set of processes}

\begin{theorem}\label{ir4}
For any integers $n$ and $f$ such that $1\leq f \leq n\!-\!2$, $\Cons(n,f\!+\!1)$ is not
	$C$-reducible to $\Cons(n,f)$.
\end{theorem}

\begin{Proof}
Suppose, for the sake of contradiction, that for some integers $n,f$ such that 
	$1\leq f \leq n-2$, we have $\Cons(n,f\!+\!1)\C \Cons(n,f)$.
We distinguish the following two cases:
\begin{enumerate}
\item $n\leq 2(f\!+\!1)$.\\
The task $\Cons(n,f)$ is  trivially a special case of $\Cons(n,f+1)$, and so $\Cons(n,f)$ 
	$C$-reduces to $\Cons(n,f+1)$.
By Proposition~\ref{generalization},  the task \ufT$(n,f)$ is a 
	generalization of $\Cons(n,f)$;   consequently $\Cons(n,f)$ $C$-reduces to \ufT$(n,f)$.
In turn, \ufT$(n,f)$ is  a special case of \ufT$(n,f\!+\!1)$, and so \ufT$(n,f)$ 
	$C$-reduces to \ufT$(n,f\!+\!1)$.
By transitivity of $\C$, it follows that  
	$$\Cons(n,f+1)\C (f\!+\!1)\mbox{-TAg}(n,f+1),$$
	which contradicts Theorem~\ref{ir3}.

\item $n> 2(f\!+\!1)$.\\
{\it A fortiori}  we have $n>2f$, and by Theorem~\ref{fnf},  it follows that  $\Cons(n,f)$ is
	$^cC$-reducible to \fT$(n,f)$.
This latter task is trivially $^cC$-reducible to \fT$(n,f\!+\!1)$ since it is a special case of 
	\fT$(n,f\!+\!1)$.
By transitivity of $\leq_{^cC}$, it follows that  
$$\Cons(n,f + 1 )\leq_{^cC} f\mbox{-TAg}(n,f + 1),$$
	which contradicts Corollary~\ref{corir2}.
\end{enumerate}\qed
\end{Proof}

A straightforward consequence  of Theorem~\ref{ir4} is the following.

\begin{corollary}\label{wfpi}
For any integers $n$ and $f$ such that $1\leq f \leq n- 2$, the wait-free Consensus task 
	$\Cons(n,n\!-\!1)$ is of higher 
	degree of unsolvability than the $f$-resilient Consensus task $\Cons(n,f)$ with respect to $\C$. 
\end{corollary} 

\subsection{Failure-Information reduction}

At this stage, it is worthy to compare Theorem~\ref{ir4} with a spinoff of the main results 
	by Chandra, Hadzilacos and Toueg in~\cite{CT96,CHT96} concerning failure detectors solving 		Consensus tasks.
In light of these two papers, it appears  that all the Consensus tasks with a majority of correct
	processes require the same information about failures to be solved.
Another way to say the same thing is that the weakest failure detectors for solving the various 
	tasks  $\Cons(n,1), \ldots, \Cons(n,\lceil n/2\rceil -1)$ are identical.
From this standpoint, all these Consensus tasks are thus equivalent.

This can be formalized by introducing a new notion of reduction -- quite different from 
	the  notions of reduction {\em \`a la}  Karp and {\em \`a la} Cook that we have studied 
	up to now -- which will measure  the hardness to solve a task in terms of the information 
	about failures that is required for solving the task.
Using  the notation and the definitions of the formal model of failure detectors in~\cite{CT96},
	we formally capture this notion of {\em Failure-Information} reduction in the following definition.

\begin{definition}\label{FIred}
Let  $T_1$ and $T_2$ be two tasks for a set $\Pi$ of processes. 
We say that $T_1$ is $FI$-{\em reducible} to $T_2$, and we note $T_1\FI T_2$, if
	any failure detector ${\cal D}$ which can be used to solve $T_2$ can also be used 
	to solve $T_1$.
\end{definition}
Since a task is solvable iff it is solvable using the trivial failure detector ${\cal D}_0$,\footnote{The
	trivial failure detector ${\cal D}_0$ is the function that maps each failure pattern $F$ to the 
	singleton $\{H_0\}$, where $H_0$ is the failure detector history such that for any time $t\in {\cal T}$
	and any process $p\in\Pi$, $H_0(p,t)=\emptyset$.
	In other words, ${\cal D}_0$ never suspects any process.} 
	we immediately derive  the following proposition.

\begin{proposition}
If $T_1$ $FI$-reduces to $T_2$ and $T_2$ is a solvable task, then $T_1$ is solvable.
\end{proposition}

Moreover,  $FI$-reduction is reflexive and transitive.
Consequently, as discussed in Part~I for our previous notions of reduction, it makes sense to order
	tasks with respect to their ``FI-difficulty'', that is with respect to $\FI$.

In particular, we define two tasks $T_1$ and $T_2$ to be FI-equivalent ($T_1\equiv_{FI} T_2$) 
	when $T_1\FI T_2$ and $T_2\FI T_1$.
As an immediate consequence of the reflexiveness and transitivity of $\FI$,
	the relation $\equiv_{FI}$ is an equivalence relation.

Arguing as  in the proof of Proposition I.6.2, we obtain that 
	$C$-reducibility implies $FI$-reducibility: if $T_1\C T_2$ then $T_1\FI T_2$.

As a consequence of the main results in~\cite{CT96,CHT96}, we are going to prove that for a fixed 
	set of processes,
	 all the Consensus tasks with a majority of correct processes have the same degree of 	
	 unsolvability with respect to $\FI$.
	 	 
\begin{proposition}\label{revcht}
For every integer $n$, $n\geq 3$, the tasks  $\Cons(n,1), \ldots,$ 
	\hbox{$ \Cons(n,\lceil n/2\rceil -1)$}
	are all  FI-equivalent.
\end{proposition}

\begin{Proof}
Since $\Cons(n,f\!+\!1)$ trivially generalizes $\Cons(n,f)$,  $\Cons(n,f)$ is $FI$-reducible to 				$\Cons(n,f\!+\!1)$.
In particular, we have
	\begin{equation}\label{fi1} 
	\Cons(n,1)\FI\cdots\FI \Cons(n,\lceil n/2\rceil -1).
	\end{equation}
	
Conversely, suppose that some failure detector ${\cal D}$ can be used to solve $\Cons(n,1)$.
By~\cite{CHT96}, we know that ${\cal D}$ is at least as strong\footnote{Here, we refer to the 
	partial ordering on failure detectors  defined in~\cite{CT96}.} as the failure detector 
	$\Omega$.
(Recall that $\Omega$ is the most general failure detector such  
	that eventually, all the correct processes always trust the same correct process.)
Moreover, Theorem~3 in~\cite{CT96} asserts that $\Cons(n,\lceil n/2\rceil -1)$ is solvable using 
	$\Omega$, and so using ${\cal D}$.
This shows that $\Cons(n,\lceil n/2\rceil -1)$ is $FI$-reducible to $\Cons(n,1)$,  i.e.,
	\begin{equation}\label{fi2} 
	\Cons(n,\lceil n/2\rceil -1) \FI  Cons(n,1).
	\end{equation}
From (\ref{fi1}) and (\ref{fi2}), it follows that all the tasks 
	$\Cons(n,1), \ldots,  \Cons(n,\lceil n/2\rceil -1).$ are FI-equivalent. \qed
\end{Proof}

Together with Theorem~\ref{ir4}, Proposition~\ref{revcht} shows that the $C$-hierarchy  
	is strictly finer  than the FI-hierarchy.
In other words, the minimal information about failures required to solve a task -- or equivalently 
	the weakest failure detector needed to solve it (if it exists) -- does not fully capture the
	hardness to solve the task.

\subsection{Wait-freedom and f-resiliency for a fixed resiliency degree}

We now prove that for a fixed resiliency degree, the smaller the set of processes is,
	the harder the Consensus tasks are.
In particular, the wait-free Consensus task $\Cons(f\!+\!1,f)$ is of higher unsolvability degree than 
	any non-wait free $f$-resilient Consensus task $\Cons(n,f)$ with respect to $\C$.

The proof  is by  a ``meta-reduction'' to the  result in the  previous section
	between wait-free and $f$-resilient Consensus tasks for a fixed set of processes:
	we shall show that from any hypothetical $C$-reduction from $\Cons(n,f)$ to $\Cons(n\!+\!1,f)$,
	we might construct a $C$-reduction from $\Cons(n,f)$ to $\Cons(n, f\!-\!1)$.
	
\begin{theorem}\label{ir5}
For any integers $n$ and $f$ such that $1\leq f \leq n-1$, $\Cons(n,f)$ is not 
	$C$-reducible to $\Cons(n\!+\!1,f)$.
\end{theorem}

\begin{Proof}
Suppose, for the sake of contradiction, that there is an algorithm $R^0$ for $\Cons(n,f)$
	using the oracle ${\cal O}.\Cons(n\!+\!1,f)$.
From $R^0$, we shall construct  an algorithm $R$ that also solves $\Cons(n,f)$ but using
	the oracle ${\cal O}.\Cons(n ,f-1)$, which contradicts Theorem~\ref{ir4}.

Let us recall that the sanctuary of ${\cal O}.T$ is $T$ (cf. Section~I.3.3).
We denote  
	$$\sigma^0=\Cons(n\!+\!1,f) \mbox{  and }  \sigma=\Cons(n,f-1)$$
	the sanctuaries of
	${\cal O}.\Cons(n\!+\!1,f)$ and ${\cal O}.\Cons(n,f-1)$, respectively.
Let $R$ be the algorithm using the oracle of sanctuary $\sigma$ such that,
	 for any process $p\in\{1,\cdots,n\}$, the automaton $R(p)$ coincides with $R^0(p)$.
We claim that $R$ solves $\Cons(n,f)$.

Let $\rho = <\!\! F, I, H \!\!>$ be any run of $R$.
From $H$, we construct a sequence $H^0$  of events as follows:
	$H^0$ is identical to $H$ except for events of type $(\sigma,p,t,\mbox{Q},v)$ and 
	$(\sigma,p,t,\mbox{A},d)$ which are replaced by $(\sigma_0,p,t,\mbox{Q},v)$ 
	and $(\sigma_0,p,t,\mbox{A},d)$, respectively.
In other words, $H^0$ is obtained from $H$ by just substituting $\sigma^0$ for  $\sigma$.
Let us now consider $\rho^0 = <\!\! F, I, H^0\!\!>$;  we argue that $\rho^0$ is a run of $R^0$.

Since $\rho$ satisfies R2 and R3, the run $\rho^0$ also satisfies R2 and R3 by construction of $H^0$.
For every  process $p$, the automata $R(p)$ and $R^0(p)$ are identical, and so 
	$\rho^0$ satisfies R4 as $\rho$ does.
From the definition of $\rho^0$, we have $Locked(\rho^0)= Locked(\rho)$;
	it follows that $\rho^0$ also satisfies R5 and R6. 
	
It remains to prove that $\rho^0$ satisfies R1.
We use the same notation as the one introduced in Section~I.4.2.
In particular, we have
$$F_{\sigma^0}=F\cup\{n+1\}\ \ \mbox{ and }\ \  F_{\sigma}=F.$$
Clearly, $H^0|\sigma^0$ is well-formed and compatible with $F_{\sigma^0}$ as 
	$H|\sigma$ is with $F_{\sigma}$.
The more delicate point to prove is that $H^0|\sigma^0$ is indeed a history of the 
	oracle ${\cal O}.\Cons(n\!+\!1,f)$, i.e, 
	$H^0|\sigma^0 \in {\cal O}.\Cons(n\!+\!1,f)(F_{\sigma^0})$.
For that, consider any consultation $H^0_k$ in $H^0$ of sanctuary $\sigma_0$;
	it naturally corresponds to a single consultation $H_k$ of $\sigma$ in $H$
	with the same queries and responses as in $H^0_k$.
Agreement in $H_k$ ensures agreement in $H^0_k$.
 For validity, form the query vector $\vec{W}$ for  $H^0_k$, and let 
	$\vec{V}^0$ be any extension of $\vec{W}$ in $\{0,1\}^{\{1,\cdots,n\!+\!1\}}$.
The projection $\vec{V}$ of  $\vec{V}^0$ onto $\{0,1\}^{\{1,\cdots,n\}}$ is an extension
	of  $\vec{W}$  in $\{0,1\}^{\{1,\cdots,n\}}$.
Since $\vec{W}$ is also the query vector 	in $H_k$, it follows that any decision $d$
	in $H^0_k$ -- which is also a decision value in $H_k$ -- is allowed by the 
	$\Cons_{\{1,\cdots, n\}}$ specification, that is 
	$$d\in \Cons_{\{1,\cdots, n\}} (F_{\sigma},\vec{V}).$$
By definition of the Consensus mappings (cf. Section~I.2.2), we have
	$$\Cons_{\{1,\cdots, n\}} (F_{\sigma},\vec{V}) \subseteq 
	\Cons_{\{1,\cdots, n\!+\!1\}}(F_{\sigma^0},\vec{V}^0)$$
	since $\vec{V}^0$ is an extension of $\vec{V}$.
It follows that $$d\in \Cons_{\{1,\cdots, n\!+\!1\}}(F_{\sigma^0},\vec{V}^0).$$
This shows that $H^0_k$ satisfies the $\Cons_{\{1,\cdots, n\!+\!1\}}$- validity condition.
Moreover, the number of queries in $H^0_k$ is the same as in $H_k$.
Therefore if there are at least $(n\!+\!1)\!-\!f$ queries in $H^0_k$, then there are at least 
	$= n\!-\!(f\!-\!1)$ queries in $H_k$, and  the oracle ${\cal O}.\Cons(n,f-1)$ necessarily 
	answers  in $H_k$.
It follows that any consultation in $H^0_k$ with at least  $(n\!+\!1)\!-\!f$ queries contains 
	a response to any correct process in $F_{\sigma^0}$.
	Hence, $H^0|\sigma^0$ is a history of  ${\cal O}.\Cons(n\!+\!1,f)$.

This shows that $\rho^0$ is a run of $R^0$.
As  $R^0$ solves the task $\Cons(n,f)$, if at most $f$ processes are faulty in $F$, 
	then $\rho^0$ satisfies the termination, irrevocability,
	agreement and $\Cons_{\{1,\cdots,n\}}$-validity conditions .
Since $\rho$ and $\rho^0$ are identical up to a renaming of $\sigma$ into $\sigma^0$, $\rho$ 
	also satisfies these conditions.
Therefore $R$  solves $\Cons(n,f)$ using  the oracle $\Cons(n,f-1)$,  a contradiction with 				Theorem~\ref{ir3}.\qed
\end{Proof}

\begin{corollary}\label{chjt}
For any integers $n$ and $f$ such that $1\leq f \leq n-2$, the wait-free Consensus
	task $\Cons(f\!+\!1,f)$ is not $C$-reducible to the $f$-resilient Consensus task 
	$\Cons(n,f)$.
\end{corollary}

\begin{Proof}
Let us assume, for the sake of contradiction, that $\Cons(f\!+\!1,f)$ is $C$-reducible to $\Cons(n,f)$
	for some integer $n$, $n\geq f+2$.
By repeated applications of Proposition~I.7.4  and transitivity of $C$-reduction, 
	we obtain that $\Cons(n-1,f)$ is $C$-reducible to $\Cons(f\!+\!1,f)$, and so
	$\Cons(n-1,f)$ is $C$-reducible to $\Cons(n,f)$,  which contradicts Theorem~\ref{ir5}.\qed
\end{Proof}

\subsection{Related work: reducibility and unsolvability}

At first sight, Theorem~\ref{ir5} and Corollary~\ref{chjt}  conflict with  Borowsky and Gafni's 				simulation~\cite{BG93, BGLR01}, 
	and more specifically with prior work  for Consensus tasks  
	by Lo and Hadzilacos~\cite{LH00}, and 
	by Chandra, Hadzilacos, Jayanti, and Toueg~\cite{CHJT94,CHJT04}. 
	
Recall that Borowsky and Gafni's  simulation consists in a general algorithm in the shared memory 		model which allows 
	a set of $f+1$ processes with at most $f$ crash failures to simulate any larger set of $n$ 
	processes also with at most $f$ crashes.
Its variant for  Consensus tasks~\cite{CHJT94} provides a  transformation of algorithms 
	that  solve the $f$-resilient Consensus task for $n$ processes using read/write registers into 			algorithms that solve the wait-free Consensus task for $f+1$ processes or using  registers
	also.\footnote{Actually, Chandra {\em et al.}  transformation works for any set of object types 
	including 	read/write registers.}
(The easily established unsolvability of $\Cons(f+1,f)$ therefore entails the unsolvability of 
	$\Cons(n,f)$ in the shared memory model.)

We could think to explain the discrepancy between the existence of such an algorithm  transformation 
	and  our irreducibility statement  in Corollary~\ref{chjt} by
	the fact that the message passing and the shared memory models are precisely not 
	equivalent here (a majority of processes may fail in the wait-free case).
However,  a closer look at this transformation  reveals  that this discrepancy actually 
	results from a more fundamental point which is worth being underlined.
	
Indeed  the transformation works as follows. 
Consider any algorithm for the task $\Cons(n,f)$ using registers, and let us fix a set of  $f+1$ processes.
The  instructions in the $n$ codes are distributed over the $f+1$ processes in a fair fashion
	way, and one by one.
The key point is that the cooperation between processes that is necessary for a correct execution 
	of the whole code for $\Cons(n,f)$ can be achieved by the processes themselves using 
	registers only.
Translating this transformation in terms of oracle-based reductions would require that processes 
	may access the internal mechanism of the oracle for $\Cons(n,f)$ for sharing it between them.
Basically, this is opposed to the notion of  oracles which are  closed black boxes that  cannot be 
	opened and dismantled.

The same argument explains the apparent contradiction between another prior work 
	about Consensus tasks in the shared memory model and the results 
	established in the previous section:
In~\cite{LH00},  Lo and Hadzilacos show how to convert any algorithm that solves 
	the one-resilient Consensus
	task  for $n$ processes using some set of object types ${\cal S}$  into an algorithm that
	solves the one-resilient Consensus task for $n-1$ processes  using the same set of  types 
	${\cal S}$, when $n$ is greater than 3.
That contradicts an immediate spinoff of Theorem~\ref{ir5} which  states that $\Cons(n-1,1)$ is not 
	$C$-reducible to $\Cons(n,1)$.
The techniques used in~\cite{CHJT94} and~\cite{LH00} are similar, and the schemes of 
	the two key transformations of Consensus algorithms are identical.
As a matter of fact,  Lo and Hadzilacos's transformation, like the one in~\cite{CHJT94}, 
	corresponds to no oracle-based reduction in the asynchronous message passing model. 	

At that point, one  might argue that the notion of oracle-based reducibility is too strong to 
	capture such algorithm transformations, and so is not  really useful.
However,  as in the classical theory of computation, oracles have been 
	introduced for the purpose of classifying undecidable/unsolvable problems/tasks.
Indeed, any reduction whose formal definition is a condition quantified over algorithms instead 
	of oracles,  of the type 
	\begin{quote}
	 $(*)\ T_1$ is reducible to $T_2$ if any algorithm solving $T_2$ can be ``transformed'' into some
	algorithm solving $T_1$,
	\end{quote}
	is trivial in the class of unsolvable tasks, since the above condition is tautologically satisfied 
	by any task $T_1$ when the task $T_2$ is unsolvable.
This observation may be applied to the pair of tasks $T_1=\Cons(f+1,f)$ and $T_2=\Cons(n,f)$, 
	or to $T_1=\Cons(n-1,1)$ and $T_2=\Cons(n,1)$, 
	and finally shows that the transformations in~\cite{CHJT94,LH00}, which actually lead 
	to unsolvability 	results, however correspond to no meaningful reduction, oracle-based or
	of the type $(*)$.

This discussion illustrates the difficulty in introducing significant and well-defined
	notions of reducibility relating unsolvable distributed tasks.
Above all, any such reducibility notion should correspond  to a hierarchy on distributed tasks, on the
	model of the Turing (resp. the Cook) hierarchy on problems, the solvable tasks playing the role 
	of decidable (resp. polynomial-time decidable) problems.
The oracle-based notions of reducibility that we have introduced in this paper, especially  
	the $C$- and $C^*$-reductions, give rise to non-trivial and sometimes unexpected results relating 
	diverse classical distributed tasks, and qualify as appropriate counterparts of the Turing
	and Cook reductions in the framework of distributed computing.	

\subsection*{Acknowledgments}

It is a pleasure to thank Andr\'e Schiper  for helpful questions and advice during 
	the writing of this paper and his careful reading of a preliminary version. 


\end{document}